\documentclass[twocolumn,english,notitlepage,aps,10pt,twocolumn]{revtex4-1}
\usepackage[T1]{fontenc}
\usepackage[latin9]{inputenc}
\usepackage{array}
\usepackage{booktabs}
\usepackage{mathrsfs}
\usepackage{multirow}
\usepackage{amsmath}
\usepackage{amssymb}
\usepackage{graphicx}
\usepackage{setspace}

\makeatletter

\providecommand{\tabularnewline}{\\}

\usepackage{fancyhdr,lipsum}
\usepackage[colorlinks,linkcolor=blue,anchorcolor=blue,citecolor=blue, urlcolor=blue]{hyperref}
\usepackage{times}
\usepackage{tikz}

\makeatother

\usepackage{babel}
\begin{document}
\title{Nonperturbative approach to the nonlinear photon echo of V-type system}
\author{Xue Zhang}
\address{Graduate School of China Academy of Engineering Physics, No. 10 Xibeiwang
East Road, Haidian District, Beijing, 100193, China}
\author{Hui Dong}
\email{hdong@gscaep.ac.cn}

\address{Graduate School of China Academy of Engineering Physics, No. 10 Xibeiwang
East Road, Haidian District, Beijing, 100193, China}
\begin{abstract}
The analysis of nonlinear spectroscopy, widely used to study the dynamics
and structures of condensed-phase matter, typically employs a perturbative
approach noticing the weak interaction between the laser and the matter
of interest. However, such perturbative approach is no longer applicable
once the interaction between the laser and the matter is strong. We
adapt the method of quantum dynamical evolution into the calculation
of signal and present the response formalism of the nonlinear spectroscopy
in a nonperturbative approach. In this new approach, we demonstrate
that in addition to the third-order term in the perturbative method,
the higher-order terms have essential contributions to the nonlinear
signal of the two-pulse and three-pulse photon echo (2PPE and 3PPE).
The detailed calculations are demonstrated with the example of a three-level
V-type system, which is widely used in the studies of quantum optics.
We consider the effect of the environment via a pure dephasing mechanism
with both the localized modes of each molecule and the shared modes
between molecules.
\end{abstract}
\maketitle

\section{Introduction}

Nonlinear spectroscopy is an increasingly important tool for studying
the dynamics of various condensed phase systems on the femtosecond
time scale\citep{Mukamel1995,Cho2009,Hamm2009}. The conventional
approach\citep{Mukamel1995,Domcke1997,Sung2001,Sung2003} to the theory
of nonlinear spectroscopy utilizes a perturbative expansion to the
third-order of the interaction between the external electric field
and the matter of interest. The perturbative approach has provided
analytical forms via the nonlinear-response functions\citep{Mukamel1995,Cho2009}.
The time-domain signals measured by heterodyne detection can be attributed
to the different nonlinear response functions, providing information
for the underlying dynamic processes\citep{Wuthrich1986,Ernst1987,Yan1991,Fleming1996,Joo1996}.
It has successfully explained many spectral phenomena and significantly
promoted the development of nonlinear spectroscopy, especially two-dimensional
spectroscopy\citep{Jonas2003,Brixner2005}. However, the calculation
of optical signal in terms of nonlinear response functions is no longer
applicable once the interaction between the external electric field
and the matter is stronger than the interaction between electrons
and nucleus or the external trap fields. For example, the strong laser
field used in the photodissociation processes\citep{Hammerich1992,Banin1994,Ashkenazi2007,Xu2018}
is so stronger than the trap field from the nucleus that the valence
electrons are ejected. And one needs alternative theoretical strategies
to treat such situation with strong interaction \citep{Seidner1995,Wolfseder1997,Gelin2003,Gelin2005,Faeder2001,Koch2002,Gelman2005,C.Jansen2000,C.Jansen2001,Mancal2006,Pisliakov2006,Ka2006,Brueggemann2007,Strelkov2016,Reshef2017,Guandalini2021}.

The nonlinear spectroscopic signal is typically calculated via the
Maxwell equation with the induced polarization vector $\boldsymbol{P}$,
which is directly obtained by calculating the expected value of the
polarization operator once the evolution of the wave function or the
density matrix is known. In the traditional perturbative approach\citep{Domcke1997,Sung2001,Sung2003,Mukamel1995},
the wave function and density matrix are obtained through the analytical
calculation with a perturbative treatment of the field-matter interaction.
To go beyond, the evolutions of the system and environment need to
be calculated with a nonperturbative method\citep{Seidner1995}. Here
we use a trajectory method to calculate the wave function. The evolutions
of the environmental degrees of freedom are mapped as trajectories
in the phase space with the coherent-state representation\citep{Scully1999}
widely used in the quantum optics. The dephasing factor induced by
the environment is acquired by summing all trajectories.

In current work, we develop the nonperturbative approach to calculate
the signal of the widely used photon echo spectroscopy\citep{Yan1991,Likforman1997,Boeij1998a,Jordanides1999,Agarwal2002},
which is an effective method to measure the homogeneous broadening
in the condensed phase. The critical element of photon echo spectroscopy
is to sort the echo signal from others with the phase matching mechanism
according to the perturbative approach. We first prove the phase matching
retained in our nonperturbative theoretical calculation. We then analytically
show that the photon echo signal have contributions from terms, which
is typically sorted as the higher-order terms in the standard perturbation
method. With the example of the V-type three-level system, we attain
the overall polarization of the system by calculating the evolution
of the wave functions. We obtained fifth- and seventh-order polarization
in addition to the third-order polarization, matching that in the
perturbative method. We show that neglecting the higher-order terms
will retain the results in the usual perturbative approach for the
weak interaction. However, these higher-order terms will affect the
measured dynamics in the two-dimensional spectroscopies\citep{Jonas2003,Brixner2005}
when the coupling between the system and the electric field is relatively
large.

The remainder of this paper is organized as follows. We introduce
the basic model Hamiltonian and general notations in Sec. II. The
nonperturbative procedure for calculating nonlinear spectroscopic
signals of two-pulse photon echo (2PPE) with and without the environment
is presented in Sec. III. The nonperturbative process for calculating
nonlinear spectroscopic signals of three-pulse photon echo (3PPE)
with environment is shown in Sec. IV. The comparison between the signal
calculated with perturbative and nonperturbative methods are illustrated
in Sec. V. Conclusions and remarks are presented in Sec. VI.

\section{Approximations and assumptions}

\begin{figure}[h]
\raggedright{}\includegraphics{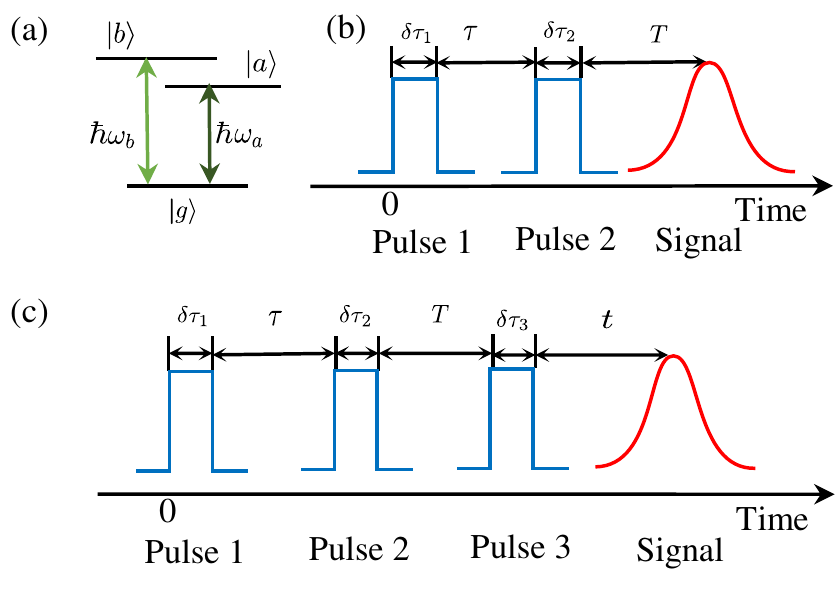}\caption{The energy level diagram (a) and the pulse sequence for (a) 2PPE and
(b) 3PPE experiments. The squares denote the incident pump pulses
under the square pulse approximation, while the dashed red lines denote
the signal.}
\label{fig:pulse_seq}
\end{figure}

We start by introducing the V-type system with the three levels denoted
as $\left|g\right\rangle $, $\left|a\right\rangle $ and $\left|b\right\rangle $.
The Hamiltonian is $H_{0}=\hbar\omega_{g}\left|g\right\rangle \left\langle g\right|+\hbar\omega_{a}\left|a\right\rangle \left\langle a\right|+\hbar\omega_{b}\left|b\right\rangle \left\langle b\right|$,
with $\omega_{b}\thickapprox\omega_{a}>\omega_{g}$. To simplify the
formula, we set the ground state energy as zero, i.e., $\omega_{g}=0$.
The energy level diagram is shown in Fig. \ref{fig:pulse_seq}(a).
The system is coupled to the laser field with the interaction as 
\begin{eqnarray}
H_{I} & = & -\left[\hbar\Omega_{a}e^{-i\nu_{a}t+i\mathbf{k}\cdot\mathbf{r}}\left|a\right\rangle \left\langle g\right|+\mathrm{h.c.}\right]\nonumber \\
 &  & -\left[\hbar\Omega_{b}e^{-i\nu_{b}t+i\mathbf{k}\cdot\mathbf{r}}\left|b\right\rangle \left\langle g\right|+\mathrm{h.c.}\right],
\end{eqnarray}
where $\nu_{a}$ and $\nu_{b}$ are the frequencies of laser field
with the wavevector $\mathbf{k}$. Here, we assume the laser field
is broad enough to cover the transition from ground state to two excited
states $\left|a\right\rangle $ and $\left|b\right\rangle $, and
consider the resonance case: $\nu_{a}=\omega_{a}$ and $\nu_{b}=\omega_{b}$.

Before derivation, we list all the approximations to be used in this
work as follows,
\begin{enumerate}
\item \textbf{Rotating-wave approximation}. It's already reflected in the
Hamiltonian. Using the rotating-wave approximation, we ignore the
high-frequency terms such as those with phase factors $e^{\pm i\left(\nu_{a}+\omega_{a}\right)t}$
and $e^{\pm i\left(\nu_{b}+\omega_{b}\right)t}$.
\item \textbf{Square pulse approximation}: We approximate the Gaussian laser
pulse as a square pulse to simplify the derivation of the laser excitation
process. The pulse duration is assumed to be $\delta\tau_{i}$ $\left(i=1,2,3\right)$.
\item \textbf{Initially system is on the ground state}. We also assume that
the system is initially on its ground state$\left|\psi\left(0\right)\right\rangle =\left|g\right\rangle $.
The light-inducing transition is around visible wavelength in the
biological system with the typical energy gap of $1.55\textrm{eV}$
(corresponding to $800\textrm{nm}$ light) between the molecular ground
and excited states. At room temperature, $k_{B}T\sim0.026\mathrm{eV}$
is so smaller than the energy gap that the molecules has no essential
population on the excited state.
\item \textbf{Initially environment on the thermal state. }In the later
discussion with system-environment interaction, we assume environmental
degrees of freedom are initially on the thermal equilibrium state,
$\rho_{\mathrm{env}}\left(0\right)=\exp\left[-\beta H_{\mathrm{env}}\right]/Z$,
with $Z=\mathrm{Tr}\left[\exp(-\beta H_{\mathrm{env}})\right]$. Here,
the detail form of Hamiltonian $H_{\mathrm{env}}$ for the environment
will be shown in the later discussion.
\end{enumerate}
The pulse sequences for the 2PPE and 3PPE experiments are illustrated
in Fig. \ref{fig:pulse_seq}(b) and (c). For simplification, we use
the notation $\theta_{i}=\Omega_{i}\delta\tau_{i}$, where $\delta\tau_{i}$
is the duration of the $i$-th pulse ($i=1$, $2$, and $3$). The
evolution operators for system under the laser pulses are denoted
as $U_{i}^{L}\left(\delta\tau_{i}\right)$, while the free evolution
operators are denoted as $U^{0}\left(x\right),$with $x=\tau,$ $T$,
and $t$. The exact expressions for the the evolution operator under
laser pulse are given in Appendix \ref{sec:Evolution-of-three-level}.

\section{Two-pulse photon echo of V-type atoms}

The photon echo experiments, including 2PPE and 3PPE, have the distinct
advantage of separating homogeneous broadening from inhomogeneous
broadening. To show how the nonperturbative approach works, we first
investigate the 2PPE, a particular case of the 3PPE, for a V-type
system without any vibrational or environmental degrees of freedom.

Assuming that the laser pulses propagate with the wave vectors $\boldsymbol{\mathbf{k}}_{1}$
and $\boldsymbol{\mathbf{k}}_{2}$ respectively, the signal is directly
observed along the $-\mathbf{k}_{1}+2\mathbf{k}_{2}$ direction in
the 2PPE experiment according to the nonperturbative approach. This
phase matching direction will be shown to retain in the nonperturbative
method in the following discussion.

The state of the system after the two pulses with a delay $\tau$
and the signal detection time $T$ can be written as

\begin{equation}
\left|\psi_{2}\left(T,\tau\right)\right\rangle =U_{0}\left(T\right)U_{2}^{L}\left(\delta\tau_{2}\right)U_{0}\left(\tau\right)U_{1}^{L}\left(\delta\tau_{1}\right)\left|g\right\rangle .
\end{equation}
The exact expression is shown with Eq. (\ref{eq:twopulsewavefunction})
in Appendix \ref{sec:State}. The echo emission is obtained as the
expect value of the transition dipole operator,

\begin{equation}
P\left(T,\tau\right)\propto\left\langle \psi_{2}\left(T,\tau\right)\right|\vec{\mu}\left|\psi_{2}\left(T,\tau\right)\right\rangle .
\end{equation}
The transition dipole has the form with the assumption of $\mu_{ag}=\mu_{bg}=\mu$,

\begin{equation}
\vec{\mu}=\mu\left|a\right\rangle \left\langle g\right|+\mu\left|b\right\rangle \left\langle g\right|+\mathrm{h.c.}
\end{equation}
We further simplify the notation by defining $s_{i}=\sin\theta_{i}$,
$c_{i}=\cos\theta_{i}$, $\eta_{a}=\Omega_{a}/\Omega$, and $\eta_{b}=\Omega_{b}/\Omega$.
In any evolution operator, the wave vector $\boldsymbol{\mathbf{k}}$
is always presented in the form of a phase factor $e^{\pm i\mathbf{k}\cdot\mathbf{r}}$
as shown in Eq. (\ref{eq:qevolution}) of Appendix. Once the interaction
between the laser pulse and the matter induces one transition, a phase
factor $e^{\pm i\mathbf{k}\cdot\mathbf{r}}$, similar to that of the
perturbative approach, is added. Therefore, the phase matching mechanism
remains in the nonperturbative approach.

With the phase matching condition, we select the echo terms with the
phase factor $\exp[i(2\mathbf{k}_{2}-\mathbf{k}_{1})\cdot\mathbf{r}]$,

\begin{eqnarray}
 &  & -i\mu s_{1}c_{1}s_{2}^{2}e^{i(2\mathbf{k}_{2}-\mathbf{k}_{1})\cdot\mathbf{r}}\left[\eta_{a}^{3}e^{-i\omega_{a}\left(T-\tau\right)}+\eta_{b}^{3}e^{-i\omega_{b}\left(T-\tau\right)}\right.\nonumber \\
 &  & \qquad\qquad\left.+\eta_{a}^{2}\eta_{b}e^{-i\omega_{b}T+i\omega_{a}\tau}+\eta_{a}\eta_{b}^{2}e^{-i\omega_{a}T+i\omega_{b}\tau}\right].\label{eq:twopulsephasematching}
\end{eqnarray}
The terms on the first line correspond to the photon echo, and the
terms on the second line do not have the rephrasing ability if $\omega_{a}$
and $\omega_{b}$ are not linearly dependent. We remark here that
the atoms or molecules are assumed to be spatially fixed with $\mathbf{r}$
independent of time. Indeed, the thermal motion of atoms and molecules
could induce a blur on the directional emission, yet it is not essential
to break the phase matching mechanism\citep{Scully1968}.

\begin{figure}
\begin{centering}
\includegraphics[viewport=0bp 0bp 142bp 160bp]{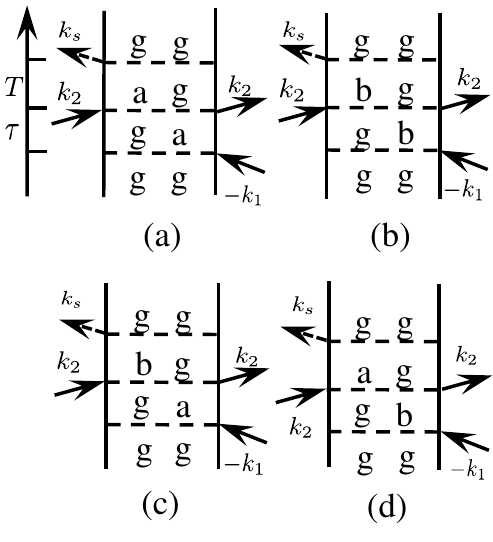}
\par\end{centering}
\caption{Double-side Feynman diagrams of 2PPE. The diagram (a), (b), (c) and
(d) corresponding to the first, second, third and last term in Eq.
(\ref{eq:twopulsephasematching}). The solid and dashed arrows in
the four diagrams represent the incident probe pulses and the signals,
respectively.}

\label{fig:Double-Feynman Diagrams of twopulse}
\end{figure}

This result is same with that obtained by the perturbative method,
except for the coefficient $s_{1}c_{1}s_{2}^{2}$ indicating the order
of terms. Therefore, we do not need to worry about the influence of
higher-order terms in the 2PPE experiment. And the perturbative method
is enough to analyze the properties of the system. The corresponding
double-side Feynman diagrams is illustrated in Fig. \ref{fig:Double-Feynman Diagrams of twopulse}. 

We will further consider the case with dephasing coupling to the environmental
degrees of freedom as 
\begin{equation}
H=H_{g}\left|g\right\rangle \left\langle g\right|+H_{a}\left|a\right\rangle \left\langle a\right|+H_{b}\left|b\right\rangle \left\langle b\right|,
\end{equation}
where 
\begin{eqnarray}
H_{g} & = & \sum\hbar\omega_{\xi}a_{\xi}^{\dagger}a_{\xi}+\sum\hbar\nu_{\varsigma}b_{\varsigma}^{\dagger}b_{\varsigma}+\sum\hbar\mu_{\zeta}c_{\zeta}^{\dagger}c_{\zeta},\\
H_{a} & = & \sum\hbar\omega_{\xi}[a_{\xi}^{\dagger}a_{\xi}+w_{\xi}(a_{\xi}^{\dagger}+a_{\xi})]+\sum\hbar\nu_{\varsigma}b_{\varsigma}^{\dagger}b_{\varsigma}\nonumber \\
 &  & +\sum\hbar\mu_{\zeta}[c_{\zeta}^{\dagger}c_{\zeta}+u_{a\zeta}(c_{\zeta}^{\dagger}+c_{\zeta})],\\
H_{b} & = & \sum\hbar\omega_{\xi}a_{\xi}^{\dagger}a_{\xi}+\sum\hbar\nu_{\varsigma}[b_{\varsigma}^{\dagger}b_{\varsigma}+v_{\varsigma}(b_{\varsigma}^{\dagger}+b_{\varsigma})]\nonumber \\
 &  & +\sum\hbar\mu_{\zeta}[c_{\zeta}^{\dagger}c_{\zeta}+u_{b\zeta}(c_{\zeta}^{\dagger}+c_{\zeta})].
\end{eqnarray}
Here, we assume the energy levels $\left|a\right\rangle $ and $\left|b\right\rangle $
have their own local environmental modes $a_{\xi}$ and $b_{\varsigma}$.
And the two levels also share common modes $c_{\zeta}$, which could
be the inter-molecular vibrations. Typically, we only consider the
local environmental modes in the discussion of energy transfer between
molecules in the condensed phase. However, such a shared mode may
exist in the strongly coupled system\citep{Jang2008,Jang2009,Hennebicq2009}.
The environmental Hamiltonian $H_{\mathrm{env}}$ is explicitly written
as $H_{\mathrm{env}}=H_{g}=\sum\hbar\omega_{\xi}a_{\xi}^{\dagger}a_{\xi}+\sum\hbar\nu_{\varsigma}b_{\varsigma}^{\dagger}b_{\varsigma}+\sum\hbar\mu_{\zeta}c_{\zeta}^{\dagger}c_{\zeta}$.

As stated in Sec. II, we assume these environmental modes are initially
on the thermal equilibrium states, $\rho_{\mathrm{env}}\left(0\right)=\exp\left[-\beta H_{\mathrm{env}}\right]/Z$.
Instead of Fock space, we will use the coherent state representation
to simplify the derivations\citep{Scully1999}. The density matrix
of the environment is written as
\begin{align}
\rho_{\mathrm{env}} & =\bigotimes_{\xi\varsigma\zeta}\int d^{2}\alpha_{\xi}d^{2}\beta_{\varsigma}d^{2}\chi_{\zeta}p(\alpha_{\xi},\beta_{\varsigma},\chi_{\zeta})\nonumber \\
 & \qquad\qquad\qquad\qquad\left|\alpha_{\xi}\beta_{\varsigma}\chi_{\zeta}\right\rangle \left\langle \alpha_{\xi}\beta_{\varsigma}\chi_{\zeta}\right|,
\end{align}
where $\left|\alpha_{\xi}\right\rangle $, $\left|\beta_{\varsigma}\right\rangle ,$
and $\left|\chi_{\zeta}\right\rangle $ are the coherent state of
the harmonic oscillator modes $a_{\xi}$, $g_{\chi}$, and $c_{\zeta}$,
respectively. The distribution is 
\begin{equation}
p(\alpha_{\xi},\beta_{\varsigma},\chi_{\zeta})=\prod_{\xi\varsigma\zeta}\frac{\exp\left[-\frac{|\alpha_{\xi}|^{2}}{n(\omega_{\xi})}-\frac{|\beta_{\varsigma}|^{2}}{n(\nu_{\varsigma})}-\frac{|\chi_{\zeta}|^{2}}{n(\mu_{\zeta})}\right]}{\pi^{3}n(\omega_{\xi})n(\nu_{\varsigma})n(\mu_{\zeta})},\label{eq:distribution}
\end{equation}
where $n\left(\omega\right)$ is mean occupation number for the state
with frequency $\omega$, i.e, $n(\omega_{\xi})=\mathrm{Tr}[a_{\xi}^{\dagger}a_{\xi}\rho_{\mathrm{env}}]$,
$n(\nu_{\varsigma})=\mathrm{Tr}[b_{\varsigma}^{\dagger}b_{\varsigma}\rho_{\mathrm{env}}]$
and $n(\mu_{\zeta})=\mathrm{Tr}[c_{\zeta}^{\dagger}c_{\zeta}\rho_{\mathrm{env}}].$
We next consider the evolution of one arbitrary state $\left|g\right\rangle \otimes\left|\alpha\beta\chi\right\rangle ,$
with corresponding subscripts ignored to simplify the notation. The
state of the system plus the environment after two pulses is shown
with Eq.(\ref{eq:twopulsewavefunctionwithenvironment}) in Appendix
\ref{sec:State}. The changes caused by the environment are reflected
in the processes of free evolution.

\begin{figure}
\raggedright{}\includegraphics[viewport=0bp 15bp 241bp 113bp]{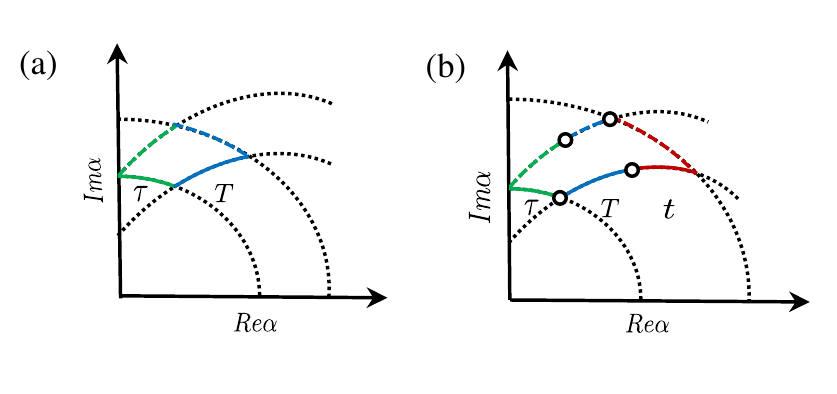}\caption{The evolution of the coherent state along two trajectories. (a) An
example of coherent state $\left|\alpha\right\rangle $ evolution
in 2PPE with evolution time $\tau$ indicated by the green line and
$T$ indicated by the blue line. (b) An example of coherent state
$\left|\alpha\right\rangle $ evolution in 3PPE with evolution time
$\tau$ indicated by the green line, $T$ indicated by the blue line,
and $t$ indicated by the red line. The solid and dashed lines represent
the same evolution time, but different evolution operators.}
\label{fig:dynamics}
\end{figure}

Keeping the terms with factors $\exp[i(2\mathbf{k}_{2}-\mathbf{k}_{1})\cdot\mathbf{r}]$,
we have the following contributions to the 2PPE, \begin{widetext}

\begin{eqnarray}
 &  & -i\mu s_{1}c_{1}s_{2}^{2}e^{i(2\mathbf{k}_{2}-\mathbf{k}_{1})\cdot\mathbf{r}}\times\left[\eta_{a}^{3}e^{-i\omega_{a}\left(T-\tau\right)}\left\langle \alpha\beta\chi\right|e^{iH_{a}\tau}e^{iH_{g}T}e^{-iH_{a}T}e^{-iH_{g}\tau}\left|\alpha\beta\chi\right\rangle \right.\nonumber \\
 &  & \qquad\qquad\qquad\qquad\qquad\quad+\eta_{b}^{3}e^{-i\omega_{b}\left(T-\tau\right)}\left\langle \alpha\beta\chi\right|e^{iH_{b}\tau}e^{iH_{g}T}e^{-iH_{b}T}e^{-iH_{g}\tau}\left|g\alpha\beta\chi\right\rangle \nonumber \\
 &  & \qquad\qquad\qquad\qquad\qquad\quad+\eta_{a}^{2}\eta_{b}e^{-i\omega_{b}T+i\omega_{a}\tau}\left\langle \alpha\beta\chi\right|e^{iH_{a}\tau}e^{iH_{g}T}e^{-iH_{b}T}e^{-iH_{g}\tau}\left|\alpha\beta\chi\right\rangle \nonumber \\
 &  & \qquad\qquad\qquad\qquad\qquad\;\left.+\eta_{a}\eta_{b}^{2}e^{-i\omega_{a}T+i\omega_{b}\tau}\left\langle \alpha\beta\chi\right|e^{iH_{b}\tau}e^{iH_{g}T}e^{-iH_{a}T}e^{-iH_{g}\tau}\left|\alpha\beta\chi\right\rangle \right].\label{eq:twopulsedephasing}
\end{eqnarray}
\end{widetext}The first two terms characterize the photon echo signal
with phase factor $e^{-i\omega_{a(b)}\left(T-\tau\right)}$. In the
following derivation, we present the derivations of each term for
the environmental response. Such response can be understand as the
overlap between the coherent states along two trajectories governed
by different evolution operators\citep{Dong2014}. One example is
illustrated in Fig. \ref{fig:dynamics}(a), we consider the overlap
between the coherent state $\left|\alpha\right\rangle $ of the first
term in Eq. (\ref{eq:twopulsedephasing}), i.e $\left\langle \alpha\right|e^{iH_{a}\tau}e^{iH_{g}T}e^{-iH_{a}T}e^{-iH_{g}\tau}\left|\alpha\right\rangle $$\rightarrow$
$\left\langle \alpha\right|A\left(-\tau\right)B\left(-T\right)A\left(T\right)B\left(\tau\right)\left|\alpha\right\rangle $
with the definition of the evolution operators $A\left(\tau\right)\equiv e^{-i\left(a^{\dagger}a+w(a^{\dagger}+a)\right)\tau}$,
and $B\left(T\right)\equiv e^{-ia^{\dagger}aT}$. The subscription
and summation symbols are ignored here. There are two evolutions of
the coherent state, $A\left(T\right)B\left(\tau\right)\left|\alpha\right\rangle $
and $B\left(T\right)A\left(\tau\right)\left|\alpha\right\rangle $.
The solid (dashed) green line represents the evolution of $B\left(\tau\right)$
($A\left(\tau\right)$), and the solid (dashed) blue line represents
the evolution of $A\left(T\right)$ ($B\left(T\right)$).

Here, the dephasing effect is captured by the trajectory factor, 
\begin{equation}
\mathcal{D}_{ij}\left(\tau,T;\alpha\beta\chi\right)=\left\langle \alpha\beta\chi\right|e^{iH_{i}\tau}e^{iH_{g}T}e^{-iH_{j}T}e^{-iH_{g}\tau}\left|\alpha\beta\chi\right\rangle ,\label{eq:trajectory factor}
\end{equation}
where $i,j=a,b$. The trajectory factor $\mathcal{D}_{ij}\left(\tau,T;\alpha\beta\chi\right)$
can be evaluated by the overlap between the wave functions of the
two evolutions. Such dephasing is caused purely by the evolution of
the environment. To measure the environmental dephasing, one must
to eliminate another dephasing caused by the ensemble average $\omega_{i}$
of the phase factors, i.e., $\exp\left[-i\omega_{i}\left(T-\tau\right)\right]$,
$\left(i=a,b\right)$ in the first two terms in Eq. (\ref{eq:twopulsedephasing})
and $\exp\left[-i\omega_{i}T+i\omega_{j}\tau\right]$ in the last
two terms in Eq. (\ref{eq:twopulsedephasing}). In the photon echo
technique, by setting $\tau=T$, the disappearance of the phase factor
$\exp\left[-i\omega_{i}\left(T-\tau\right)\right]$ in the first two
terms of Eq. (\ref{eq:twopulsedephasing}) allows the direct observation
of dephasing caused by its coupling to environment modes without any
additional dephasing induced by averaging over distribution of transition
frequencies (inhomogeneous broadening). The last two terms show the
dephasing between two excited states $\left|a\right\rangle $ , $\left|b\right\rangle $
and ground state $\left|g\right\rangle $. The inhomogeneity induced
dephasing through the phase term $\exp\left[-i\omega_{i}T+i\omega_{j}\tau\right]$
$\left(i,j=a,b;\;i\neq j\right)$ can not be eliminated through the
photon echo if $\omega_{i}$ and $\omega_{j}$ are not linearly dependent
on each other.

The total dephasing factor $D\left(\tau,T\right)$ as

\begin{equation}
D(\tau,T)=\prod_{\xi\varsigma\zeta}\int d^{2}\alpha_{\xi}d^{2}\beta_{\varsigma}d^{2}\chi_{\zeta}p(\alpha_{\xi},\beta_{\varsigma},\chi_{\zeta})\mathcal{D}_{ij}(\tau,T;\alpha\beta\chi),
\end{equation}
is easily calculated through Eq. (\ref{eq:distribution}) and Eq.
(\ref{eq:trajectory factor}). We take the first term $\mathcal{D}_{aa}\left(\tau,T;\alpha\beta\chi\right)$
as an example and derive the total dephasing factor $D_{aa}\left(\tau,T\right)$.
The exact form of $\mathcal{D}_{aa}\left(\tau,T;\alpha\beta\chi\right)$
is presented with Eq. (\ref{eq:D_aa_trajectoryfactor}) in Appendix
\ref{sec:Trajectory-dephasing-factor}. A simple multivariate Gaussian
integration gives the follow result, 
\begin{eqnarray}
 &  & D_{aa}(\tau,T)=\exp\left\{ -2g_{a}^{*}(\tau)-g_{a}^{*}(T)-g_{a}(T)+g_{a}^{*}(\tau+T)\right.\nonumber \\
 &  & \quad\quad\left.-2\mathscr{C}_{aa}^{*}(\tau)-\mathscr{C}_{aa}^{*}(T)-\mathscr{C}_{aa}(T)+\mathscr{C}_{aa}^{*}(\tau+T)\right\} ,
\end{eqnarray}
where 
\begin{align}
g_{a}\left(t,T_{e}\right) & =\int d\omega\mathcal{J}_{a}\left(\omega\right)\left[\left(1-\cos\omega t\right)\coth(\omega/2kT_{e})\right.\nonumber \\
 & \qquad\qquad\qquad\left.+i\left(-\omega t+\sin\omega t\right)\right],\\
\mathscr{C}_{aa}\left(t,T_{e}\right) & =\int d\mu\mathcal{J}_{aa}\left(\mu\right)\left[\left(1-\cos\mu t\right)\coth(\nu/2kT_{e})\right.\nonumber \\
 & \qquad\qquad\qquad\left.+i\left(-\nu t+\sin\nu t\right)\right].
\end{align}
Here $g_{a}\left(t,T_{e}\right)$ and $\mathscr{C}_{aa}\left(t,T_{e}\right)$
are the lineshape functions at temperature $T_{e}$. The spectral
density is defined as $\mathcal{J}_{a}\left(\omega\right)=\sum w_{\xi}^{2}\delta(\omega-\omega_{\xi})$
and $\mathcal{J}_{aa}\left(\mu\right)=\sum u_{a\varsigma}^{2}\delta(\mu-\mu_{\varsigma})$.

At the echo time $T=\tau$, the intensity will have the following
form
\begin{equation}
S\left(\tau\right)=S\left(0\right)\Gamma_{aa}'\left(\tau\right),
\end{equation}
where 
\begin{align}
 & \Gamma_{aa}'\left(\tau\right)\equiv|D_{aa}\left(\tau,\tau\right)|^{2}\nonumber \\
 & \qquad=\exp\left[-2(4P_{a}(\tau)+4P_{c}(\tau)-P_{a}(2\tau)-P_{c}(2\tau))\right],\label{eq:echo1}
\end{align}
with $P_{a}\left(t\right)=\mathrm{Re}[g_{a}\left(t\right)]$, $P_{c}\left(t\right)=\mathrm{Re}[\mathscr{C}_{aa}\left(t\right)]$.
Until here, we have derived the well-known photon echo formalism for
two-level system ($\left|g\right\rangle $,$\left|a\right\rangle $).
And Eq. \ref{eq:echo1} clearly shows the photon echo signal intensity
is determined only by system-environment coupling.

For the V-type system, we have another term

\begin{align}
D_{ab}\left(\tau,T\right) & =\exp[-g_{a}^{*}(\tau)-g_{b}(T)-\mathscr{C}_{aa}^{*}(\tau)-\mathscr{C}_{bb}(T)]\nonumber \\
 & \times\exp[-\mathscr{C}_{ab}^{*}(\tau)-\mathscr{C}_{ab}^{*}(T)+\mathscr{C}_{ab}^{*}(\tau+T)],
\end{align}
where

\begin{align}
\mathscr{C}_{ab}\left(t,T_{e}\right) & =\int d\mu\mathcal{J}_{ab}\left(\mu\right)\left[\left(1-\cos\mu t\right)\coth(\nu/2kT_{e})\right.\nonumber \\
 & \qquad\qquad\qquad\left.+i\left(-\nu t+\sin\nu t\right)\right],
\end{align}
and the spectral density is defined as $\mathcal{J}_{ab}\left(\mu\right)=\sum u_{a\varsigma}u_{b\varsigma}\delta(\mu-\mu_{\varsigma})$.
Besides the dephasing of $\left|g\right\rangle \left\langle a\right|$
during the waiting time $\tau$ and $\left|b\right\rangle \left\langle g\right|$
during the waiting time $T$ with the term $\exp[-g_{a}^{*}(\tau)-g_{b}(T)]$,
we have additional dephasing due to the shared modes for the two excited
states represented by the term $\exp[-\mathscr{C}_{aa}^{*}(\tau)-\mathscr{C}_{ab}^{*}(\tau)-\mathscr{C}_{ab}^{*}(T)-\mathscr{C}_{bb}(T)+\mathscr{C}_{ab}^{*}(\tau+T)]$.

However, the observed coherence for the delay time $T$ and $\tau$
is always between the ground state ($\left|g\right\rangle $) and
excited states ($\left|a\right\rangle $ and $\left|b\right\rangle $)
illustrated by the Feynman diagrams in Fig. \ref{fig:Double-Feynman Diagrams of twopulse},
with no access to the coherence between two excited states. To get
more dynamics information about the three-level system, we will study
the 3PPE and show the accessibility of observation of coherence between
two excited states in the next section.

\section{Three-pulse photon echo of V-type atoms\label{sec:Three-pulse-echo-of}}

Sec. II demonstrates that the nonperturbative method is equivalent
to the perturbative method in 2PPE spectroscopy. To show the difference
between the two methods, we then study the 3PPE, a typical way to
measure the homogeneous broadening.

Assuming that the three pulses propagates with the wave vectors $\boldsymbol{\mathbf{k}}_{1}$
, $\boldsymbol{\mathbf{k}}_{2}$ and $\boldsymbol{\mathbf{k}}_{3}$,
respectively, the signal is directly observed along the $\boldsymbol{-\boldsymbol{\mathbf{k}}_{1}+\boldsymbol{\mathbf{k}}_{2}+\boldsymbol{\mathbf{k}}_{3}}$
direction in the 3PPE experiment. And the phase matching direction
is the same as that of the perturbative method. Here, the derivation
for the system without coupling to the environment is skipped. We
directly consider the V-type system with coupling to the environment.
The state of the system after three pulses is 
\begin{align}
\left|\psi_{3}\left(\tau,T,t\right)\right\rangle  & =U_{0}\left(t\right)U_{3}^{L}\left(\delta\tau_{3}\right)U_{0}\left(T\right)U_{2}^{L}\left(\delta\tau_{2}\right)\nonumber \\
 & \qquad\qquad\qquad U_{0}\left(\tau\right)U_{1}^{L}\left(\delta\tau_{1}\right)\left|g\alpha\beta\chi\right\rangle ,
\end{align}
whose exact expression is shown with Eq. (\ref{eq:threepulsewavefunctionwithenvironment})
in Appendix \ref{sec:State}.

Following a similar derivation procedure in 2PPE, we keep the terms
with the phase factor $\exp\left[i\left(-\mathbf{k}_{1}+\mathbf{k}_{2}+\mathbf{k}_{3}\right)\cdot\mathbf{r}\right]$.
It has 20 terms in total, including 8 third-order terms, 8 fifth-order
terms, and 4 seventh-order terms. Here, by the number of the order,
we follow the language of the perturbative methods with the assumption
of weak interaction between the system and the three laser fields.
Previously, the response function formalism was established based
on the perturbative theory, and the higher-order (larger than third-order)
terms are ignored. Compared to the traditional perturbative approach,
these 20 terms corresponding to photon echo are grouped according
to their order and listed as follows.

\begin{figure*}
\begin{raggedright}
\includegraphics{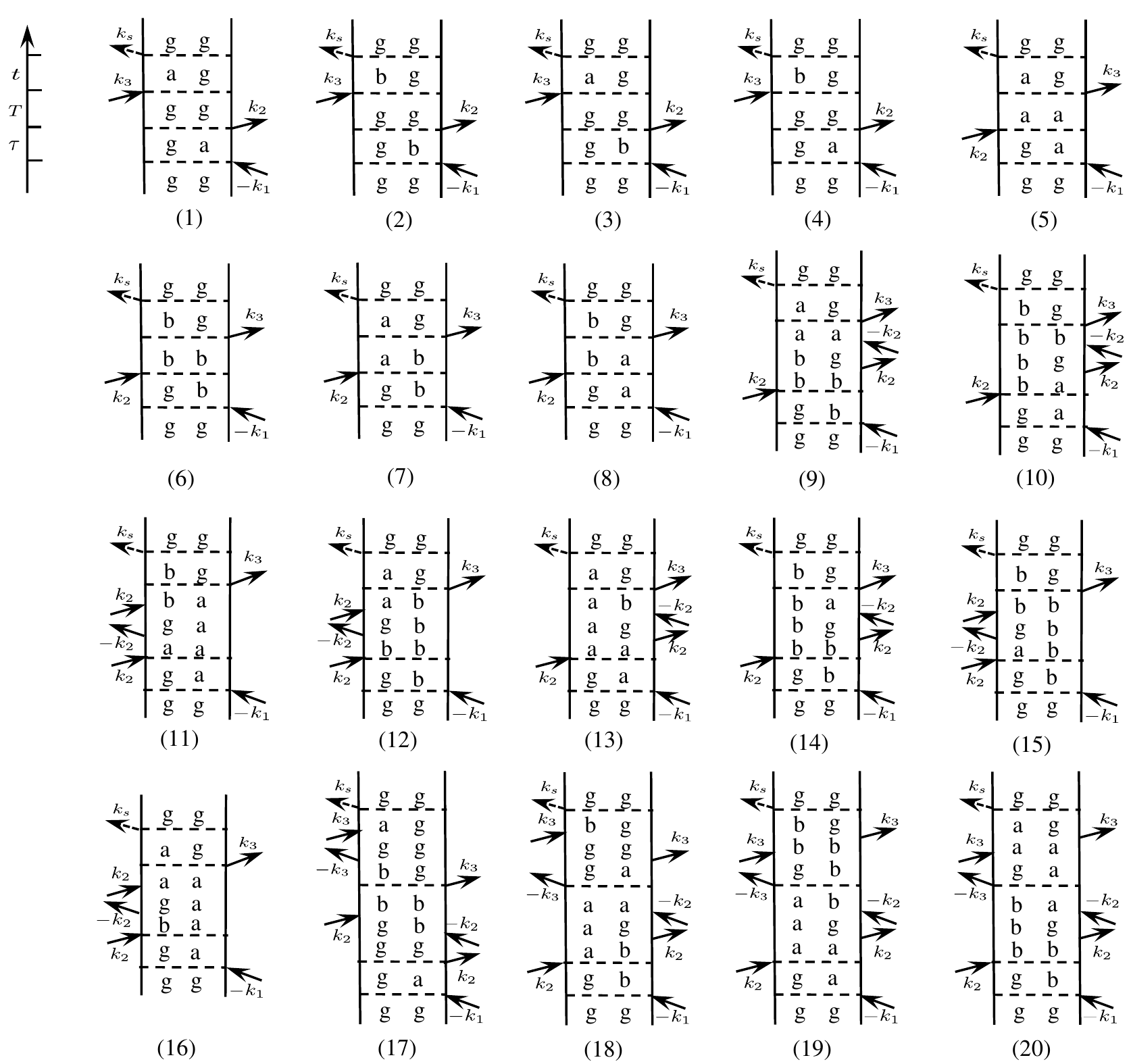}
\par\end{raggedright}
\caption{Double-side Feynman diagrams for 20 terms in 3PPE. (1)-(8) third-order
terms in Eq. (\ref{eq:third-order terms}) , (9)-(16) fifth-order
terms in Eq. (\ref{eq:fifth-order terms}) and (17)-(20) seventh-order
terms in Eq. (\ref{eq:seventh-order terms}). The sequence numbers
(1)-(20) are consistent with that in 20 terms. The solid and dashed
arrows in the three diagrams are the same as Fig. \ref{fig:Double-Feynman Diagrams of twopulse}.}
\label{fig:Double-Feynman Diagrams of threepulse}
\end{figure*}

\begin{widetext}

\textbf{Third-order Terms: }(The factor $\exp\left[i\left(-\mathbf{k}_{1}+\mathbf{k}_{2}+\mathbf{k}_{3}\right)\cdot\mathbf{r}\right]$
are ignored in the following formula. Here, $\Delta_{ab}=\omega_{b}-\omega_{a}$
is the energy spacing between the state $\left|a\right\rangle $ and
$\left|b\right\rangle $.)

\begin{eqnarray}
\mathbf{1} & \qquad & -i\mu\eta_{a}^{3}s_{1}c_{1}s_{2}c_{2}s_{3}c_{3}e^{i\omega_{a}\left(\tau-t\right)}\left\langle \alpha\beta\chi\right|e^{iH_{a}\tau}e^{iH_{g}\left(T+t\right)}e^{-iH_{a}t}e^{-iH_{g}\left(T+\tau\right)}\left|\alpha\beta\chi\right\rangle ,\nonumber \\
\mathbf{2} & \qquad & -i\mu\eta_{b}^{3}s_{1}c_{1}s_{2}c_{2}s_{3}c_{3}e^{i\omega_{b}\left(\tau-t\right)}\left\langle \alpha\beta\chi\right|e^{iH_{b}\tau}e^{iH_{g}\left(T+t\right)}e^{-iH_{b}t}e^{-iH_{g}\left(T+\tau\right)}\left|\alpha\beta\chi\right\rangle ,\nonumber \\
\mathbf{3} & \qquad & -i\mu\eta_{a}\eta_{b}^{2}s_{1}c_{1}s_{2}c_{2}s_{3}c_{3}e^{i\omega_{b}\tau-i\omega_{a}t}\left\langle \alpha\beta\chi\right|e^{iH_{b}\tau}e^{iH_{g}\left(T+t\right)}e^{-iH_{a}t}e^{-iH_{g}\left(T+\tau\right)}\left|\alpha\beta\chi\right\rangle ,\nonumber \\
\mathbf{4} & \qquad & -i\mu\eta_{a}^{2}\eta_{b}s_{1}c_{1}s_{2}c_{2}s_{3}c_{3}e^{i\omega_{a}\tau-i\omega_{b}t}\left\langle \alpha\beta\chi\right|e^{iH_{a}\tau}e^{iH_{g}\left(T+t\right)}e^{-iH_{b}t}e^{-iH_{g}\left(T+\tau\right)}\left|\alpha\beta\chi\right\rangle ,\nonumber \\
\mathbf{5} & \qquad & -i\mu\eta_{a}^{3}s_{1}c_{1}s_{2}s_{3}(\eta_{b}^{2}+\eta_{a}^{2}c_{2})(\eta_{b}^{2}+\eta_{a}^{2}c_{3})e^{i\omega_{a}\left(\tau-t\right)}\left\langle \alpha\beta\chi\right|e^{iH_{a}\left(\tau+T\right)}e^{iH_{g}t}e^{-iH_{a}\left(t+T\right)}e^{-iH_{g}\tau}\left|\alpha\beta\chi\right\rangle ,\nonumber \\
\mathbf{6} & \qquad & -i\mu\eta_{b}^{3}s_{1}c_{1}s_{2}s_{3}(\eta_{a}^{2}+\eta_{b}^{2}c_{2})(\eta_{a}^{2}+\eta_{b}^{2}c_{3})e^{i\omega_{b}\left(\tau-t\right)}\left\langle \alpha\beta\chi\right|e^{iH_{b}\left(\tau+T\right)}e^{iH_{g}t}e^{-iH_{b}\left(t+T\right)}e^{-iH_{g}\tau}\left|\alpha\beta\chi\right\rangle ,\nonumber \\
\mathbf{7} & \qquad & -i\mu\eta_{a}\eta_{b}^{2}s_{1}c_{1}s_{2}s_{3}(\eta_{a}^{2}+\eta_{b}^{2}c_{2})(\eta_{b}^{2}+\eta_{a}^{2}c_{3})e^{i\omega_{b}\tau-i\omega_{a}t+i\Delta_{ab}T}\left\langle \alpha\beta\chi\right|e^{iH_{b}\left(\tau+T\right)}e^{iH_{g}t}e^{-iH_{a}\left(t+T\right)}e^{-iH_{g}\tau}\left|\alpha\beta\chi\right\rangle ,\nonumber \\
\mathbf{8} & \qquad & -i\mu\eta_{a}^{2}\eta_{b}s_{1}c_{1}s_{2}s_{3}(\eta_{b}^{2}+\eta_{a}^{2}c_{2})(\eta_{a}^{2}+\eta_{b}^{2}c_{3})e^{i\omega_{a}\tau-i\omega_{b}t-i\Delta_{ab}T}\left\langle \alpha\beta\chi\right|e^{iH_{a}\left(\tau+T\right)}e^{iH_{g}t}e^{-iH_{b}\left(t+T\right)}e^{-iH_{g}\tau}\left|\alpha\beta\chi\right\rangle .\label{eq:third-order terms}
\end{eqnarray}

\textbf{Fifth-order Terms: }
\begin{eqnarray}
\mathbf{9} & \qquad & -i\mu\eta_{a}^{3}\eta_{b}^{2}s_{1}c_{1}s_{2}(c_{2}-1)s_{3}(\eta_{b}^{2}+\eta_{a}^{2}c_{3})e^{i\omega_{b}\tau-i\omega_{a}t}\left\langle \alpha\beta\chi\right|e^{iH_{b}\tau}e^{iH_{a}T}e^{iH_{g}t}e^{-iH_{a}\left(t+T\right)}e^{-iH_{g}\tau}\left|\alpha\beta\chi\right\rangle ,\nonumber \\
\mathbf{10} & \qquad & -i\mu\eta_{a}^{2}\eta_{b}^{3}s_{1}c_{1}s_{2}(c_{2}-1)s_{3}(\eta_{a}^{2}+\eta_{b}^{2}c_{3})e^{i\omega_{a}\tau-i\omega_{b}t}\left\langle \alpha\beta\chi\right|e^{iH_{a}\tau}e^{iH_{b}T}e^{iH_{g}t}e^{-iH_{b}\left(t+T\right)}e^{-iH_{g}\tau}\left|\alpha\beta\chi\right\rangle ,\nonumber \\
\mathbf{11} & \qquad & -i\mu\eta_{a}^{4}\eta_{b}s_{1}c_{1}s_{2}(\eta_{b}^{2}+\eta_{a}^{2}c_{2})s_{3}(c_{3}-1)e^{i\omega_{a}\tau-i\omega_{b}t}\left\langle \alpha\beta\chi\right|e^{iH_{a}\left(\tau+T\right)}e^{iH_{g}t}e^{-iH_{b}t}e^{-iH_{a}T}e^{-iH_{g}\tau}\left|\alpha\beta\chi\right\rangle ,\nonumber \\
\mathbf{12} & \qquad & -i\mu\eta_{a}\eta_{b}^{4}s_{1}c_{1}s_{2}(\eta_{a}^{2}+\eta_{b}^{2}c_{2})s_{3}(c_{3}-1)e^{i\omega_{b}\tau-i\omega_{a}t}\left\langle \alpha\beta\chi\right|e^{iH_{b}\left(\tau+T\right)}e^{iH_{g}t}e^{-iH_{a}t}e^{-iH_{b}T}e^{-iH_{g}\tau}\left|\alpha\beta\chi\right\rangle ,\nonumber \\
\mathbf{13} & \qquad & -i\mu\eta_{a}^{3}\eta_{b}^{2}s_{1}c_{1}s_{2}(c_{2}-1)s_{3}(\eta_{b}^{2}+\eta_{a}^{2}c_{3})e^{i\omega_{a}\tau-i\omega_{a}t+i\Delta_{ab}T}\left\langle \alpha\beta\chi\right|e^{iH_{a}\tau}e^{iH_{b}T}e^{iH_{g}t}e^{-iH_{a}\left(t+T\right)}e^{-iH_{g}\tau}\left|\alpha\beta\chi\right\rangle ,\nonumber \\
\mathbf{14} & \qquad & -i\mu\eta_{a}^{2}\eta_{b}^{3}s_{1}c_{1}s_{2}(c_{2}-1)s_{3}(\eta_{a}^{2}+\eta_{b}^{2}c_{3})e^{i\omega_{b}\tau-i\omega_{b}t-i\Delta_{ab}T}\left\langle \alpha\beta\chi\right|e^{iH_{b}\tau}e^{iH_{a}T}e^{iH_{g}t}e^{-iH_{b}\left(t+T\right)}e^{-iH_{g}\tau}\left|\alpha\beta\chi\right\rangle ,\nonumber \\
\mathbf{15} & \qquad & -i\mu\eta_{a}^{2}\eta_{b}^{3}s_{1}c_{1}s_{2}(\eta_{a}^{2}+\eta_{b}^{2}c_{2})s_{3}(c_{3}-1)e^{i\omega_{b}\tau-i\omega_{b}t+i\Delta_{ab}T}\left\langle \alpha\beta\chi\right|e^{iH_{b}\left(\tau+T\right)}e^{iH_{g}t}e^{-iH_{b}t}e^{-iH_{a}T}e^{-iH_{g}\tau}\left|\alpha\beta\chi\right\rangle ,\nonumber \\
\mathbf{16} & \qquad & -i\mu\eta_{a}^{3}\eta_{b}^{2}s_{1}c_{1}s_{2}(\eta_{b}^{2}+\eta_{a}^{2}c_{2})s_{3}(c_{3}-1)e^{i\omega_{a}\tau-i\omega_{a}t-i\Delta_{ab}T}\left\langle \alpha\beta\chi\right|e^{iH_{a}\left(\tau+T\right)}e^{iH_{g}t}e^{-iH_{a}t}e^{-iH_{b}T}e^{-iH_{g}\tau}\left|\alpha\beta\chi\right\rangle .\label{eq:fifth-order terms}
\end{eqnarray}

\textbf{Seventh-order Terms:}

\begin{eqnarray}
\mathbf{17} & \qquad & -i\mu\eta_{a}^{3}\eta_{b}^{4}s_{1}c_{1}s_{2}(c_{2}-1)s_{3}(c_{3}-1)e^{i\omega_{a}\left(\tau-t\right)}\left\langle \alpha\beta\chi\right|e^{iH_{a}\tau}e^{iH_{b}T}e^{iH_{g}t}e^{-iH_{a}t}e^{-iH_{b}T}e^{-iH_{g}\tau}\left|\alpha\beta\chi\right\rangle ,\nonumber \\
\mathbf{18} & \qquad & -i\mu\eta_{a}^{4}\eta_{b}^{3}s_{1}c_{1}s_{2}(c_{2}-1)s_{3}(c_{3}-1)e^{i\omega_{b}\left(\tau-t\right)}\left\langle \alpha\beta\chi\right|e^{iH_{b}\tau}e^{iH_{a}T}e^{iH_{g}t}e^{-iH_{b}t}e^{-iH_{a}T}e^{-iH_{g}\tau}\left|\alpha\beta\chi\right\rangle ,\nonumber \\
\mathbf{19} & \qquad & -i\mu\eta_{a}^{4}\eta_{b}^{3}s_{1}c_{1}s_{2}(c_{2}-1)s_{3}(c_{3}-1)e^{i\omega_{a}\tau-i\omega_{b}t+i\Delta_{ab}T}\left\langle \alpha\beta\chi\right|e^{iH_{a}\tau}e^{iH_{b}T}e^{iH_{g}t}e^{-iH_{b}t}e^{-iH_{a}T}e^{-iH_{g}\tau}\left|\alpha\beta\chi\right\rangle ,\nonumber \\
\mathbf{20} & \qquad & -i\mu\eta_{a}^{3}\eta_{b}^{4}s_{1}c_{1}s_{2}(c_{2}-1)s_{3}(c_{3}-1)e^{i\omega_{b}\tau-i\omega_{a}t-i\Delta_{ab}T}\left\langle \alpha\beta\chi\right|e^{iH_{b}\tau}e^{iH_{a}T}e^{iH_{g}t}e^{-iH_{a}t}e^{-iH_{b}T}e^{-iH_{g}\tau}\left|\alpha\beta\chi\right\rangle .\label{eq:seventh-order terms}
\end{eqnarray}
\end{widetext}In the perturbative approach, the terms \textbf{9-20}
are not considered due to their negligible contributions to total
photon echo signal intensity. In all responses of the third-order,
terms \textbf{1-4} represent the ground state dynamics $\left|g\right\rangle \left\langle g\right|$
during the second delay time $T$, while term 5 (6) corresponds to
the dynamics of the excited state $\left|a\right\rangle \left\langle a\right|$
($\left|b\right\rangle \left\langle b\right|$). The corresponding
double-side Feynman diagrams are illustrated in Fig. \ref{fig:Double-Feynman Diagrams of threepulse}.

We next utilize two-dimensional spectroscopy, a powerful experimental
technique that probes the nonlinear optical response of materials,
to study the effect of higher-order terms on the signal. In essence,
the traditional \textquotedblleft 1D\textquotedblright{} spectroscopy,
which measures the linear response, reveals the excitations in a system,
whereas the two-dimensional spectroscopy reveals the dynamics caused
by these excitations\citep{Tokmakoff2011}. Through the two-dimensional
spectroscopy, we get the dynamics of the system and explain spectral
phenomena. 

To get the two-dimensional spectrum $S_{i}\left(\omega_{\tau},T,\omega_{t}\right)$,
we apply Fourier transform for the first and third delay times $t$
and $\tau$ to get the two-dimensional spectrum, namely, $S_{i}\left(\omega_{\tau},T,\omega_{t}\right)=\mathcal{F}\left[R_{i}\left(\tau,T,t\right)\right]$
($i$=1, 2, ..., 20). For example, the Fourier transform of term 1
results in a peak around $\left(\omega_{a},-\omega_{a}\right)$, noticing
the response term of the environment will not present a significant
change of oscillation frequencies during the first and third waiting
times $t$ and $\tau$. In the later discussion, we will flip the
sign of the second axis to be positive as $S'_{i}\left(\omega_{\tau},T,\omega_{t}\right)\equiv S_{i}\left(\omega_{\tau},T,-\omega_{t}\right)$
and give the signal as $\left|S'_{i}\left(\omega_{\tau},T,\omega_{t}\right)\right|^{2}$.

We group the 20 response terms based on their position on the two-dimensional
spectrum and show them in Table \ref{tab:position}. During the waiting
time $T,$ the coherence dynamics for signal calculated by the traditional
perturbative method at the off-diagonal peak ($\omega_{a},\omega_{b}$)
show a beating with the frequency as energy difference $\Delta_{ab}$
between two excited states, which is contributed by the term 8. However,
due to the contribution of an additional higher-order term 19, the
coherence dynamics in our nonperturbative method at the off-diagonal
peak ($\omega_{a},\omega_{b}$) exhibit different oscillations at
frequencies $\Delta_{ab}$ and $2\Delta_{ab}$. We have contributions
from higher-order terms to the dynamics of each peak.

In the traditional perturbative method, the design of the 3PPE allows
the direct probe of the dephasing between the two electronic energy
levels $\left|a\right\rangle $ and $\left|b\right\rangle $ via the
off-diagonal terms. The oscillation pattern enables the separation
of coherence dynamics from population dynamics and we can get more
information about the molecules, which is why 3PPE is commonly used
with respect to 2PPE. 

\begin{table}
\begin{spacing}{1.2}
\begin{centering}
\begin{tabular}{cccc}
\toprule 
Position & terms & order & T dynamics\tabularnewline
\midrule
\multirow{5}{*}{$\left(\omega_{a},\omega_{b}\right)$} & 4 & third-order & ground state $\left|g\right\rangle \left\langle g\right|$\tabularnewline
 & 8 & third-order & coherence $\left|b\right\rangle \left\langle a\right|$\tabularnewline
 & 10 & fifth-order & excited state $\left|b\right\rangle \left\langle b\right|$\tabularnewline
 & 11 & fifth-order & excited state $\left|a\right\rangle $$\left\langle a\right|$\tabularnewline
 & 19 & seventh-order & coherence $\left|a\right\rangle \left\langle b\right|$\tabularnewline
\midrule 
\multirow{5}{*}{$\left(\omega_{b},\omega_{a}\right)$} & 3 & third-order & ground state $\left|g\right\rangle \left\langle g\right|$\tabularnewline
 & 7 & third-order & coherence $\left|a\right\rangle \left\langle b\right|$\tabularnewline
 & 9 & fifth-order & excited state $\left|a\right\rangle $$\left\langle a\right|$\tabularnewline
 & 12 & fifth-order & excited state $\left|b\right\rangle \left\langle b\right|$\tabularnewline
 & 20 & seventh-order & coherence $\left|b\right\rangle \left\langle a\right|$\tabularnewline
\midrule 
\multirow{5}{*}{$\left(\omega_{a},\omega_{a}\right)$} & 1 & third-order & ground state $\left|g\right\rangle \left\langle g\right|$\tabularnewline
 & 5 & third-order & excited state $\left|a\right\rangle \left\langle a\right|$\tabularnewline
 & 13 & fifth-order & coherence $\left|a\right\rangle \left\langle b\right|$\tabularnewline
 & 16 & fifth-order & coherence $\left|b\right\rangle \left\langle a\right|$\tabularnewline
 & 17 & seventh-order & excited state $\left|b\right\rangle \left\langle b\right|$\tabularnewline
\midrule 
\multirow{5}{*}{$\left(\omega_{b},\omega_{b}\right)$} & 2 & third-order & ground state $\left|g\right\rangle \left\langle g\right|$\tabularnewline
 & 6 & third-order & excited state $\left|b\right\rangle \left\langle b\right|$\tabularnewline
 & 14 & fifth-order & coherence $\left|b\right\rangle \left\langle a\right|$\tabularnewline
 & 15 & fifth-order & coherence $\left|a\right\rangle \left\langle b\right|$\tabularnewline
 & 18 & seventh-order & excited state $\left|a\right\rangle $$\left\langle a\right|$\tabularnewline
\bottomrule
\end{tabular}
\par\end{centering}
\end{spacing}
\caption{Classification of 20 terms. These 20 terms are divided into 4 categories
according to their spectral positions on the two-dimensional Fourier
spectrum. We also list the corresponding order and $T$-time dynamics
of each term.}

\label{tab:position}
\end{table}

The decoherence factor for 3PPE is different from that in 2PPE, especially
for the processes with the coherence dynamics along delay time $T$.
A schematic diagram of the evolution of the coherent state $\left|\alpha\right\rangle $
in term 8 is illustrated in Fig. \ref{fig:dynamics}(b). There are
two evolutions of the coherent state, $e^{-iH_{b}\left(t+T\right)}e^{-iH_{g}\tau}\left|\alpha\right\rangle $
and $e^{-iH_{g}t}e^{-iH_{a}\left(\tau+T\right)}\left|\alpha\right\rangle $.
The red lines represents the evolution of $t$. We only consider the
environmental factor in the coherence pathway of term 8. The trajectory
decoherence factor for term 8 is $\mathcal{D}_{8}\left(\tau,T,t;\alpha,\beta,\chi\right)$,
presented in Eq. (\ref{eq:D_8_trajectoryfactor}) in Appendix \ref{sec:Trajectory-dephasing-factor}.
The integration over all the trajectories with the weight function
$p\left(\alpha_{\xi},\beta_{\varsigma},\chi_{\zeta}\right)$ results
in the decoherence factor
\begin{eqnarray}
D_{8}\left(\tau,T,t\right) & = & \exp\left[-g_{a}^{*}(T+\tau)-g_{b}(t+T)-\mathscr{C}_{aa}^{*}(T+\tau)\right.\nonumber \\
 &  & \qquad\qquad\left.-\mathscr{C}_{bb}(t+T)+\mathscr{C}_{ab}^{*}(T+t+\tau)\right.\nonumber \\
 &  & \qquad\qquad\left.-\mathscr{C}_{ab}^{*}(\tau)+\mathscr{C}_{ab}(T)-\mathscr{C}_{ab}^{*}(t)\right].
\end{eqnarray}
Decoherence factors for all terms in Sec. \ref{sec:Three-pulse-echo-of}
are presented in Appendix \ref{sec:Decoherence-Factor of three pulses}.

\section{Comparisons}

In the two sections above, we have derived both the 2PPE and 3PPE
formalism with the nonperturbative method. To compared with the traditional
perturbative approach, we focus on the 3PPE, and discuss the effect
brought by the higher-order terms.

For simplicity, we use the well-known Kubo stochastic model\citep{Kubo1969},
where the lineshape function is given by\citep{Hanamura1983}
\begin{equation}
g\left(t\right)=(\Delta\tau_{0})^{2}[e^{-t/\tau_{0}}+\frac{t}{\tau_{0}}-1].\label{eq:lineshape_function}
\end{equation}
Here $\tau_{0}$ is the environmental correlation time, and $\Delta$
is the fluctuation amplitude of the excited states induced by the
coupling to the environment. To clarify the advantage of our nonperturbative
approach, we first consider the case where levels $\left|a\right\rangle $
and $\left|b\right\rangle $ don't have shared environmental modes
$c_{\zeta}$, namely $\mathscr{C}_{aa}\left(t\right)=0$ , $\mathscr{C}_{bb}\left(t\right)=0$
and $\mathscr{C}_{ab}\left(t\right)=0$. And we also assume the equal
coupling strength $w_{\xi}=v_{\varsigma}$, which implies $g_{a}\left(t\right)=g_{b}\left(t\right)=g\left(t\right)$.
The correlation time and fluctuation amplitude are chosen as $\tau_{0}=1\textrm{ps}$
and $\Delta=15\textrm{cm}^{-1}$, respectively. The pulse durations
are $35\textrm{fs}$. The energy level gaps of the two excited states
are $\omega_{a}=1.25\times10^{4}\textrm{cm}^{-1}$ and $\omega_{b}=1.31\times10^{4}\textrm{cm}^{-1}$,
respectively.

\begin{figure*}
\includegraphics{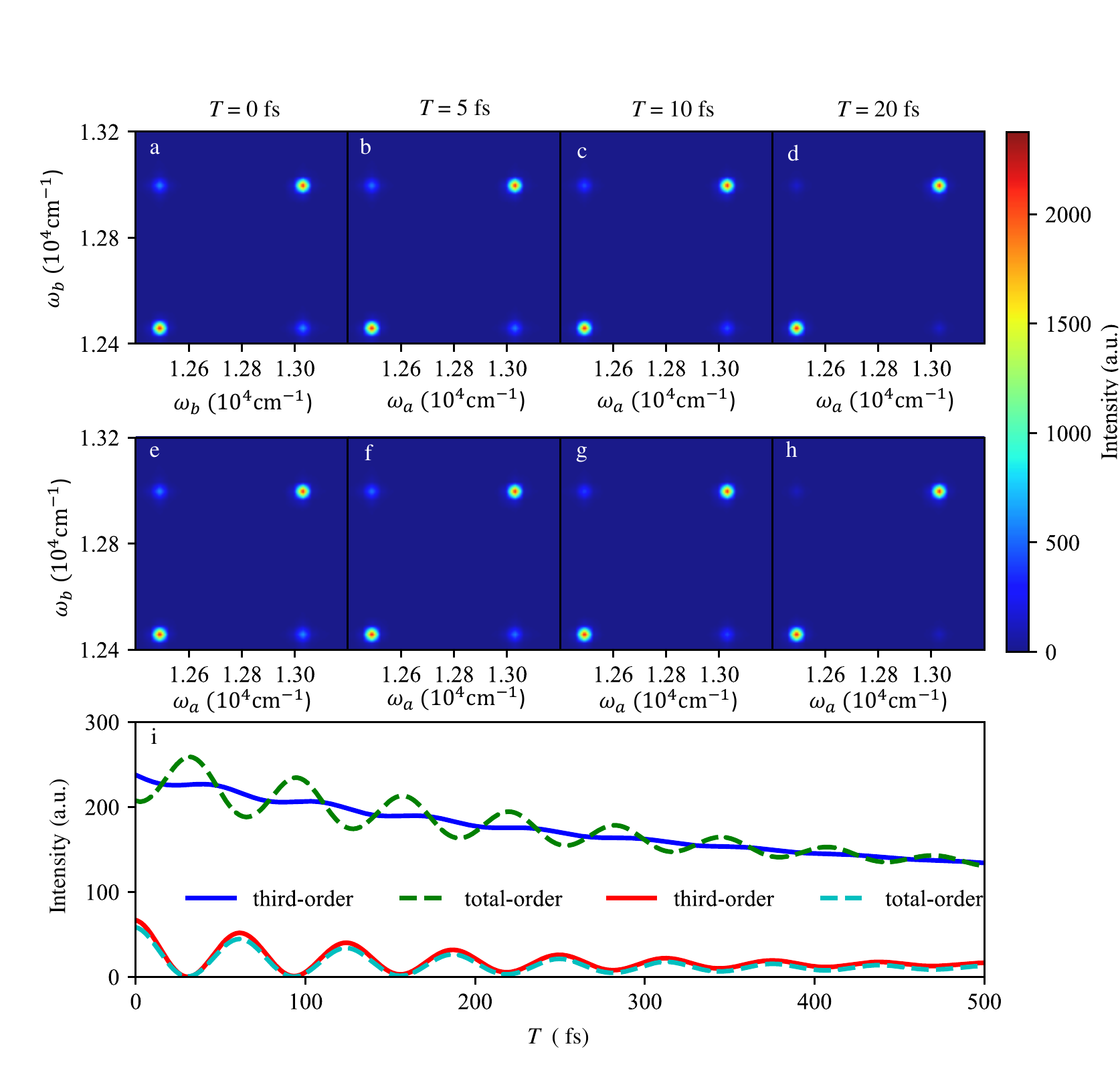}

\caption{(Color online) 2D spectra and the dynamics along the waiting time
$T$ for the weak interaction. (a)-(d) the third-order terms and (e)-(f)
the total-order terms for $T=0,5,10,20\thinspace\textrm{fs}$,respectively.
(i) the dynamics along the waiting time $T$ for the diagonal $\left(\omega_{a},\omega_{a}\right)$
and off-diagonal $\left(\omega_{a},\omega_{b}\right)$ peaks for third-order
terms and total-order terms. The solid blue and dashed green lines
describe the information of the diagonal peak, and the solid red line
and dashed cyan lines describe the information of the off-diagonal
peak. These spectra for any fixed $T$ are obtained via 2D fast Fourier
transform of the time domain signal scanned from 0 to 5 ps with 0.5fs
time step for both $\tau$ and $t$. In the figure, we plot the absolute
values of the transform result. And the coupling strengths between
system and electric field are $\Omega_{a}=53\textrm{cm}^{-1}$ and
$\Omega_{b}=53\textrm{cm}^{-1}$.}
\label{fig:signal}
\end{figure*}

Fig. \ref{fig:signal} shows the 3PPE signal of the V-type system
with relatively small coupling strengths $\Omega_{a}=53\textrm{cm}^{-1}$
and $\Omega_{b}=53\textrm{cm}^{-1}$. The first row (a-d) show the
signal calculated from the conventional perturbative approach and
the second row (e-h) show the signal obtained with our nonperturbative
approach. Through the vertical comparison of (a) and (e), (b) and
(f), (c) and (g), (d) and (h), we find that the spectral features
are roughly the same for both two methods. We also compare the dynamics
along the waiting time $T$ in Fig. \ref{fig:signal}(i). The blue
and red lines are the evolution of diagonal peak $\left(\omega_{a},\omega_{a}\right)$
and off-diagonal peak $\left(\omega_{a},\omega_{b}\right)$ for the
third-order terms; meanwhile, the green and cyan lines are the evolution
of diagonal peak $\left(\omega_{a},\omega_{a}\right)$ and off-diagonal
peak $\left(\omega_{a},\omega_{b}\right)$ for the total-order terms.
The slight oscillation of the solid blue line is caused by the discrete
Fourier transform. It's clear that the dynamics along the waiting
time $T$ are basically the same for both the diagonal and off-diagonal
peaks. The oscillation of the green line comes from the fifth-order
coherence terms 13 and 16, yet is relatively small. The red and cyan
lines are almost identical, indicating the same information containing
coherence terms that cause vibrations, and the coherence brought about
by higher-order terms can be ignored.

\begin{figure*}
\includegraphics{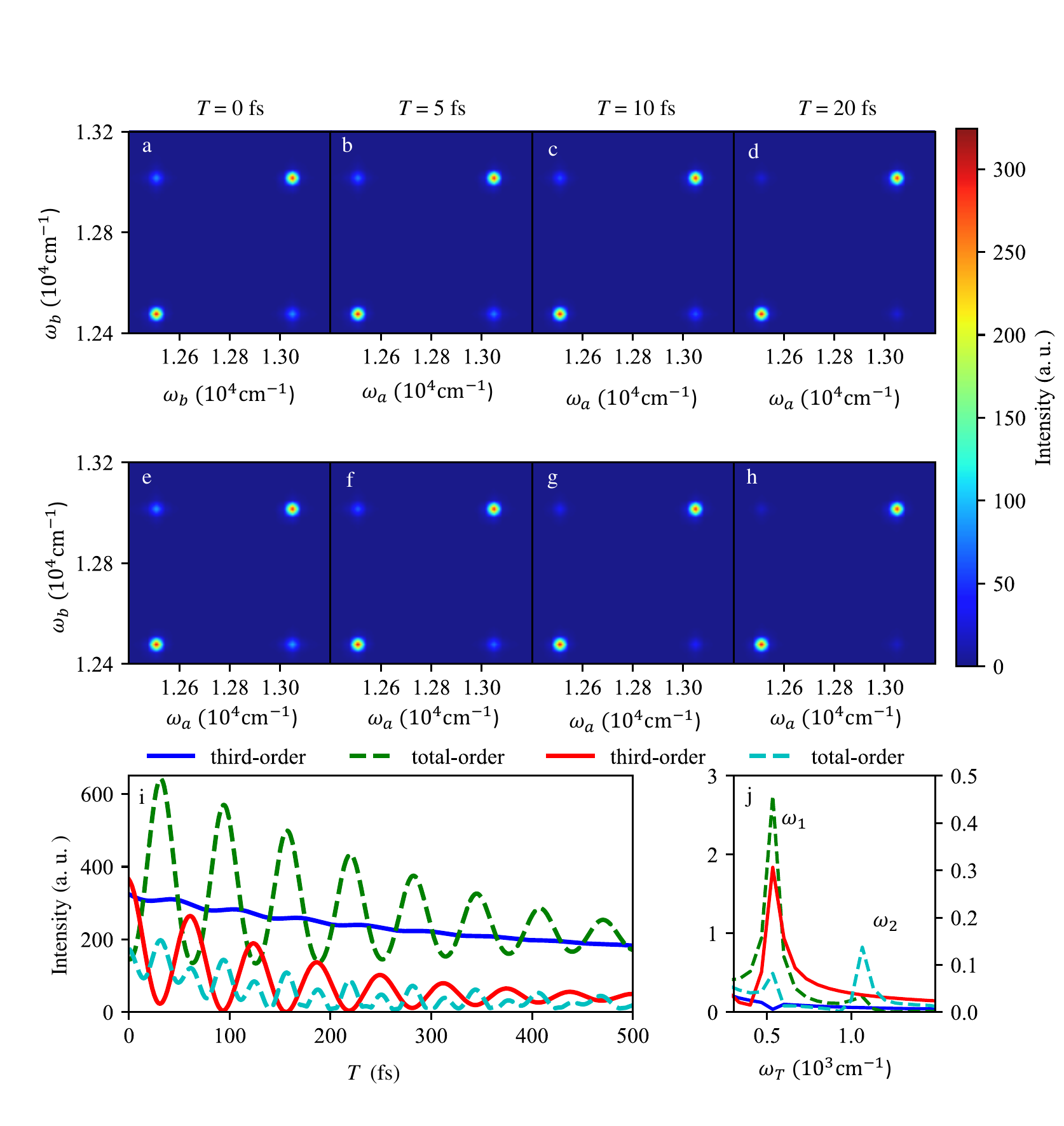}

\caption{(Color online) 2D spectra and the dynamics along the waiting time
$T$ for the strong interaction. (a)-(d) the third-order terms and
(e)-(f) the total-order terms for $T=0,5,10,20\thinspace\textrm{fs}$,respectively.
The coupling strengths between system and electric field are $\Omega_{a}=106\textrm{cm}^{-1}$
and $\Omega_{b}=106\textrm{cm}^{-1}$. (i) the dynamics along the
waiting time $T$ for the diagonal $\left(\omega_{a},\omega_{a}\right)$
and off-diagonal $\left(\omega_{a},\omega_{b}\right)$ peaks for third-order
terms and total-order terms. (j) the Fast Fourier Transform shows
the frequency of oscillation of each line in (i).}
\label{fig:signal-1}
\end{figure*}

Now, we turn to the case with strong couplings. In Eq. (\ref{eq:third-order terms}),
Eq. (\ref{eq:fifth-order terms}) and Eq. (\ref{eq:seventh-order terms}),
the coupling strengths between the laser and the matter only affect
the formula coefficients $s_{i}$ and $c_{i}$, which vary periodically
with the coupling strength. Therefore, we only consider relatively
strong coupling strengths $\Omega\delta\tau\sim\pi/4$. In Fig. \ref{fig:signal-1},
the coupling strengths are $\Omega_{a}=106\textrm{cm}^{-1}$ and $\Omega_{b}=106\textrm{cm}^{-1}$.
We find that the spectral features are roughly the same in Fig. \ref{fig:signal-1}(a)-(h)
for both two methods. However, the distinctly different evolution
over waiting time T is presented in Fig. \ref{fig:signal-1}(i) between
perturbative and nonperturbative methods. The green line has a larger
additional oscillation than the blue line, and the cyan line also
has different oscillation compared to the red line. To learn more
about these oscillations, we perform discrete Fourier transforms of
the signal in Fig. \ref{fig:signal-1}(i) scanned from 0 to 500 fs
with 0.5fs time step for time $T$, the result is shown in Fig. \ref{fig:signal-1}(j).
To compare the effects of higher-order terms on the signals of the
diagonal and off-diagonal peaks, respectively, the blue solid and
green dashed lines share the left y-axis, and the red solid and cyan
dashed lines share the right y-axis. There are two two frequencies
$\omega_{1}\sim\Delta_{ab}$ and $\omega_{2}\sim2\Delta_{ab}$. The
two oscillation frequencies of the green line are the result of the
contribution of terms 13 and 16 to the signal. The oscillation frequency
$\omega_{1}$ in the red line is the result of the contribution of
the term 8 to the signal. Due to the contribution of the term 19 in
the higher-order terms to the signal, the cyan line has an oscillation
$\omega_{2}$ compared to the red line. Clearly, the high-order terms
will modify the dynamics presented in the signal, and cannot be ignored
in strong interaction.

To clarify the influence of the shared modes, we next consider the
case where levels $\left|a\right\rangle $ and $\left|b\right\rangle $
have shared environmental modes $c_{\zeta}$ with the same coupling
strength $u_{a\zeta}=u_{b\zeta}=u_{\zeta}$, namely $\mathscr{C}_{aa}\left(t\right)=\mathscr{C}_{bb}\left(t\right)=\mathscr{C}_{ab}\left(t\right)$.
Here, we still use the Kubo model with the lineshape function Eq.
(\ref{eq:lineshape_function}). And we assume the equal coupling strength
$w_{\xi}=v_{\varsigma}=u_{\zeta}$, which implies the parameters in
lineshape function are identical for the three environmental modes.

\begin{figure*}
\includegraphics{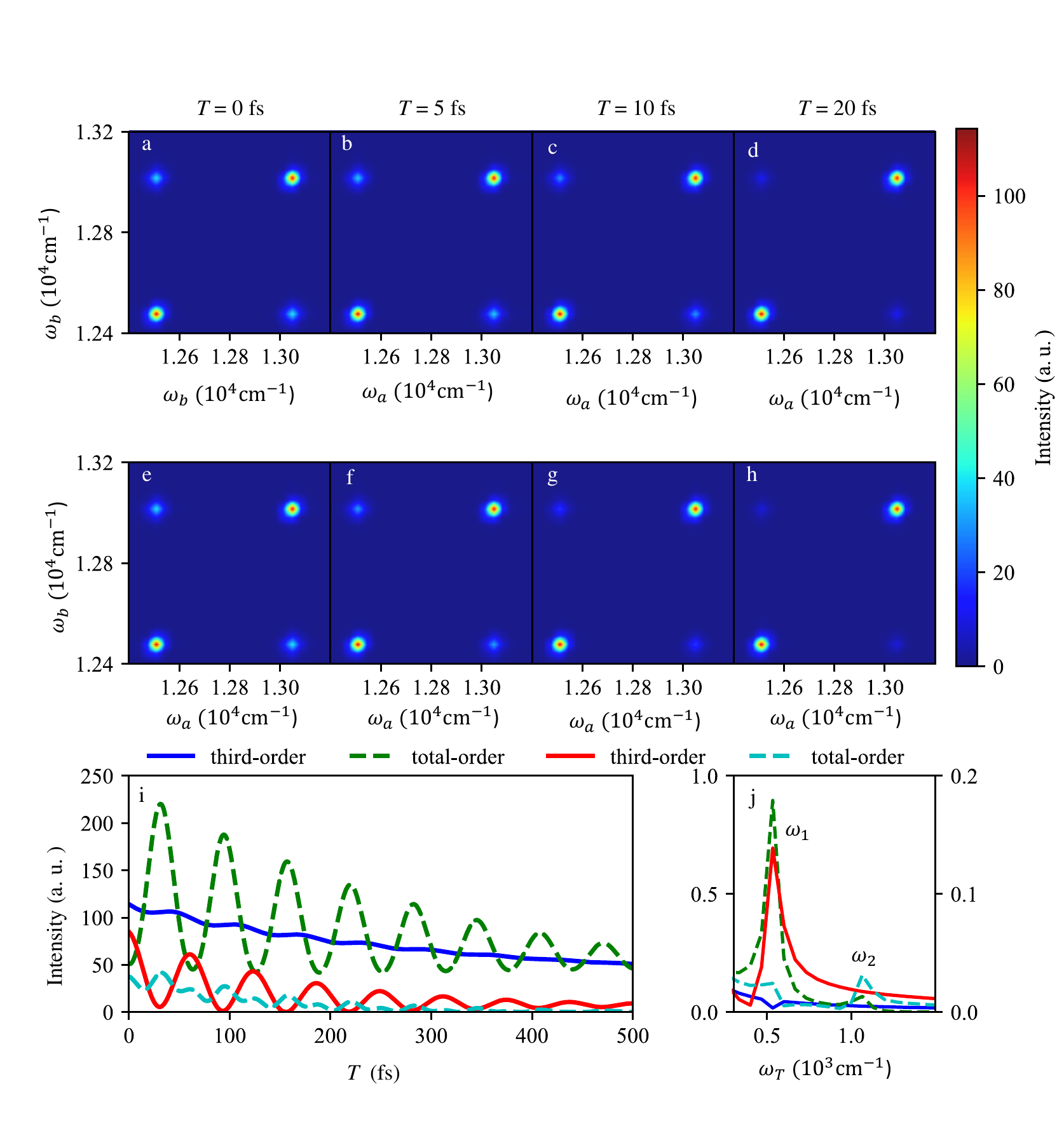}

\caption{(Color online) 2D spectra and the dynamics along the waiting time
$T$ for the strong interaction containing shared modes $c_{j}$.
The parameters are the same as in Fig. \ref{fig:signal-1}.}
\label{fig:signal-2}
\end{figure*}

We consider the same coupling strengths with Fig. \ref{fig:signal-1},
and the results are presented in Fig. \ref{fig:signal-2}. The spectral
features are roughly the same according to the 2D Fourier transform
spectra (a)-(h) in Fig. \ref{fig:signal-2} and \ref{fig:signal-1}.
Actually, the dynamics along the waiting time $T$ in Fig. \ref{fig:signal-2}(i)
are same with that in Fig. \ref{fig:signal-1}(i). And the oscillation
frequency of the signal reflected in Fig. \ref{fig:signal-2}(j) are
also similar. It indicates that the shared modes only affect the amplitude
of the signal without causing additional decoherence.

\section{Conclusions and remarks}

In this paper, we have theoretically derived the nonlinear responses
of 2PPE and 3PPE using a nonperturbative approach, starting from the
Hamiltonian with and without environment, and obtained the analytical
expressions of the response functions based on a three-level V-type
system. 

In 2PPE, our methods are same with the traditional methods, with only
a slight difference in amplitude. In 3PPE, the expressions have additional
fifth- and seventh-order terms compared to that in the perturbation
method. In order to explore the effect of these higher-order terms.
We then utilize the Kubo model and find that in the case of strong
coupling, the nonperturbative approach shows additional features in
the signal. During the dynamics at time $T$, the oscillatory behaviors
of the diagonal and off-diagonal peaks has changed greatly, and these
changes come from the contributions of higher-order terms. However,
in the case of weak coupling, the changes of the oscillatory behaviors
are not obvious. Our nonperturbative approach is more suitable than
the perturbative method for strong coupling.

Finally, we further examine the impact of the shared mode. We found
that the shared mode only affects the amplitude of the signal without
causing additional decoherence in the dynamics along time $T$.

\bibliographystyle{apsrev4-1}
\bibliography{nonperturbative}

\appendix
\begin{widetext}

\section{Evolution of three-level system under laser pulse\label{sec:Evolution-of-three-level}}

The three levels are denoted as $g$, $a$ and $b$. The Hamiltonian
of the three-level V-type system is $H_{0}=\hbar\omega_{g}\left|g\right\rangle \left\langle g\right|+\hbar\omega_{a}\left|a\right\rangle \left\langle a\right|+\hbar\omega_{b}\left|b\right\rangle \left\langle b\right|,$
with $\omega_{b}\approx\omega_{a}>\omega_{g}$. To simplify the formula,
we set $\omega_{g}=0$. Interacting with the laser pulse, the system
is coupled to the laser field 
\begin{eqnarray}
H_{I} & = & -\left[\hbar\Omega_{a}e^{-i\nu_{a}t+i\mathbf{k}\cdot\mathbf{r}}\left|a\right\rangle \left\langle g\right|+\mathrm{h.c.}\right]\nonumber \\
 &  & -\left[\hbar\Omega_{b}e^{-i\nu_{b}t+i\mathbf{k}\cdot\mathbf{r}}\left|b\right\rangle \left\langle g\right|+\mathrm{h.c.}\right],
\end{eqnarray}
where $\nu$ is the frequency of laser field with the wavevector $\mathbf{k}$.
And the In the interaction picture, the Hamiltonian would be time-independent,
\begin{eqnarray}
H_{I}^{\mathrm{int}} & = & -\left[\hbar\Omega_{a}e^{i\mathbf{k}\cdot\mathbf{r}}\left|a\right\rangle \left\langle g\right|+\mathrm{h.c.}\right]\nonumber \\
 &  & -\left[\hbar\Omega_{b}e^{i\mathbf{k}\cdot\mathbf{r}}\left|b\right\rangle \left\langle g\right|+\mathrm{h.c.}\right].
\end{eqnarray}
The general solution of the current system under the interaction with
laser pulse can be presented by the state as 
\[
\left|\Psi(t)\right\rangle =C_{g}(t)\left|g\right\rangle +C_{a}(t)\left|a\right\rangle +C_{b}(t)\left|b\right\rangle ,
\]
where 
\begin{eqnarray}
C_{g}(t) & = & C_{g}(0)\cos\Omega t+i(\frac{\Omega_{a}e^{-i\mathbf{k}\cdot\mathbf{r}}}{\Omega}C_{a}(0)+\frac{\Omega_{b}e^{-i\mathbf{k}\cdot\mathbf{r}}}{\Omega}C_{b}(0))\sin\Omega t,\nonumber \\
C_{a}(t) & = & i\frac{\Omega_{a}e^{i\mathbf{k}\cdot\mathbf{r}}}{\Omega}\sin\Omega tC_{g}(0)+\frac{\Omega_{b}^{2}+\Omega_{a}^{2}\cos\Omega t}{\Omega^{2}}C_{a}(0)+\frac{[\cos\Omega t-1]\Omega_{a}\Omega_{b}}{\Omega^{2}}C_{b}(0).\nonumber \\
C_{b}(t) & = & i\frac{\Omega_{b}e^{i\mathbf{k}\cdot\mathbf{r}}}{\Omega}\sin\Omega tC_{g}(0)+\frac{[\cos\Omega t-1]\Omega_{a}\Omega_{b}}{\Omega^{2}}C_{a}(0)+\frac{\Omega_{a}^{2}+\Omega_{b}^{2}\cos\Omega t}{\Omega^{2}}C_{b}(0).\label{eq:qevolution}
\end{eqnarray}
For the special cases that is initially on the ground state $C_{g}(0)=1$
to be used in the derivation, we have

\begin{eqnarray}
C_{g}(t) & = & \cos\Omega t,\\
C_{a}(t) & = & i\frac{\Omega_{a}e^{i\mathbf{k}\cdot\mathbf{r}}}{\Omega}\sin\Omega t,\\
C_{b}(t) & = & i\frac{\Omega_{b}e^{i\mathbf{k}\cdot\mathbf{r}}}{\Omega}\sin\Omega t.
\end{eqnarray}

\section{The state after evolution\label{sec:State}}

We first give the exact expression of the state in 2PPE. The state
of the system without environment after the two pulses with a delay
$\tau$ can be written as

\begin{eqnarray}
 &  & \left|\psi_{2}\left(T,\tau\right)\right\rangle \nonumber \\
 & = & \left[\cos\theta_{1}\cos\theta_{2}-e^{i\left(\mathbf{k}_{1}-\mathbf{k}_{2}\right)\cdot\mathbf{r}}\sin\theta_{1}\sin\theta_{2}\left(\frac{\Omega_{a}^{2}}{\Omega^{2}}e^{-i\omega_{a}\tau}+\frac{\Omega_{b}^{2}}{\Omega^{2}}e^{-i\omega_{b}\tau}\right)\right]\left|g\right\rangle \nonumber \\
 &  & +i\cos\theta_{1}\sin\theta_{2}\frac{\Omega_{a}e^{i\mathbf{k}_{2}\cdot\mathbf{r}-i\omega_{a}T}}{\Omega}\left|a\right\rangle \nonumber \\
 &  & +i\frac{\sin\theta_{1}e^{i\mathbf{k}_{1}\cdot\mathbf{r}-i\omega_{a}T}}{\Omega^{3}}\left(\Omega_{a}\left(\Omega_{b}^{2}+\Omega_{a}^{2}\cos\theta_{2}\right)e^{-i\omega_{a}\tau}+[\cos\theta_{2}-1]\Omega_{a}^{2}\Omega_{b}e^{-i\omega_{b}\tau}\right)\left|a\right\rangle \nonumber \\
 &  & +i\cos\theta_{1}\sin\theta_{2}\frac{\Omega_{b}e^{i\mathbf{k}_{2}\cdot\mathbf{r}-i\omega_{b}T}}{\Omega}\left|b\right\rangle \nonumber \\
 &  & +i\frac{\sin\theta_{1}e^{i\mathbf{k}_{1}\cdot\mathbf{r}-i\omega_{b}T}}{\Omega^{3}}\left([\cos\theta_{2}-1]\Omega_{a}^{2}\Omega_{b}e^{-i\omega_{a}\tau}+\Omega_{b}\left(\Omega_{a}^{2}+\Omega_{b}^{2}\cos\theta_{2}\right)e^{-i\omega_{b}\tau}\right)\left|b\right\rangle .\label{eq:twopulsewavefunction}
\end{eqnarray}

The state of the system coupled with environment after the two pulses
with a delay $\tau$ can be written as

\begin{eqnarray}
 &  & \left|\psi_{2}\left(\tau,T\right)\right\rangle \nonumber \\
 & = & \left[c_{1}c_{2}e^{-iH_{g}\left(T+\tau\right)}-\eta_{a}^{2}s_{1}s_{2}e^{i\left(\mathbf{k}_{1}-\mathbf{k}_{2}\right)\cdot\mathbf{r}-i\omega_{a}\tau}e^{-iH_{g}T}e^{-iH_{a}\tau}-\eta_{b}^{2}s_{1}s_{2}e^{i\left(\mathbf{k}_{1}-\mathbf{k}_{2}\right)\cdot\mathbf{r}-i\omega_{b}\tau}e^{-iH_{g}T}e^{-iH_{b}\tau}\right]\left|g\alpha\beta\chi\right\rangle \label{eq:twopulsewavefunctionwithenvironment}\\
 &  & +\left[i\eta_{a}c_{1}s_{2}e^{i\mathbf{k}_{2}\cdot\mathbf{r}-i\omega_{a}T}e^{-iH_{a}T}e^{-iH_{g}\tau}+i\eta_{a}s_{1}\left(\eta_{b}^{2}+\eta_{a}^{2}c_{2}\right)e^{i\mathbf{k}_{1}\cdot\mathbf{r}-i\omega_{a}\tau-i\omega_{a}T}e^{-iH_{a}T}e^{-iH_{a}\tau}\right.\nonumber \\
 &  & \qquad\qquad\qquad\qquad\qquad\qquad\qquad\qquad\qquad\left.+i\eta_{b}s_{1}\left(c_{2}-1\right)\eta_{a}\eta_{b}e^{i\mathbf{k}_{1}\cdot\mathbf{r}-i\omega_{b}\tau-i\omega_{a}T}e^{-iH_{a}T}e^{-iH_{b}\tau}\right]\left|a\alpha\beta\chi\right\rangle \nonumber \\
 &  & +\left[i\eta_{b}c_{1}s_{2}e^{i\mathbf{k}_{2}\cdot\mathbf{r}-i\omega_{b}T}e^{-iH_{b}T}e^{-iH_{g}\tau}+i\eta_{a}s_{1}\left(c_{2}-1\right)\eta_{a}\eta_{b}e^{i\mathbf{k}_{1}\cdot\mathbf{r}-i\omega_{a}\tau-i\omega_{b}T}e^{-iH_{b}T}e^{-iH_{a}\tau}\right.\nonumber \\
 &  & \qquad\qquad\qquad\qquad\qquad\qquad\qquad\qquad\qquad\left.+i\eta_{b}s_{1}\left(\eta_{a}^{2}+\eta_{b}^{2}c_{2}\right)e^{i\mathbf{k}_{1}\cdot\mathbf{r}-i\omega_{b}\tau-i\omega_{b}T}e^{-iH_{b}T}e^{-iH_{b}\tau}\right]\left|b\alpha\beta\chi\right\rangle .
\end{eqnarray}

We then give the exact expression of the state in 3PPE. The state
of the system coupled with environment after the three pulses with
a delay $\tau$ and $T$ can be written as

\begin{eqnarray}
 &  & \left|\psi_{3}\left(\tau,T,t\right)\right\rangle \nonumber \\
 & = & \left[c_{1}\left|g\right\rangle e^{-iH_{g}t}+i\eta_{a}s_{3}\left|a\right\rangle e^{i\mathbf{k}_{3}\cdot\mathbf{r}-i\omega_{a}t-iH_{a}t}+i\eta_{b}s_{3}\left|b\right\rangle e^{i\mathbf{k}_{3}\cdot\mathbf{r}-i\omega_{b}t-iH_{b}t}\right]\nonumber \\
 &  & \otimes\left[c_{1}c_{2}e^{-iH_{g}\left(T+\tau\right)}-\eta_{a}^{2}s_{1}s_{2}e^{i\left(\mathbf{k}_{1}-\mathbf{k}_{2}\right)\cdot\mathbf{r}-i\omega_{a}\tau}e^{-iH_{g}T}e^{-iH_{a}\tau}\right.\nonumber \\
 &  & \qquad\qquad\qquad\qquad\qquad\:\quad-\left.\eta_{b}^{2}s_{1}s_{2}e^{i\left(\mathbf{k}_{1}-\mathbf{k}_{2}\right)\cdot\mathbf{r}-i\omega_{b}\tau}e^{-iH_{g}T}e^{-iH_{b}\tau}\right]\left|\alpha\beta\chi\right\rangle \nonumber \\
 &  & +\left[i\eta_{a}s_{3}\left|g\right\rangle e^{-i\mathbf{k}_{3}\cdot\mathbf{r}-iH_{g}t}+(\eta_{b}^{2}+\eta_{a}^{2}c_{3})\left|a\right\rangle e^{-i\omega_{a}t-iH_{a}t}+(c_{3}-1)\eta_{a}\eta_{b}\left|b\right\rangle e^{-i\omega_{b}t-iH_{b}t}\right]\nonumber \\
 &  & \quad\otimes\left[i\eta_{a}c_{1}s_{2}e^{i\mathbf{k}_{2}\cdot\mathbf{r}-i\omega_{a}T}e^{-iH_{a}T}e^{-iH_{g}\tau}+i\eta_{a}s_{1}(\eta_{b}^{2}+\eta_{a}^{2}c_{2})e^{i\mathbf{k}_{1}\cdot\mathbf{r}-i\omega_{a}\left(\tau+T\right)}e^{-iH_{a}\left(\tau+T\right)}\right.\nonumber \\
 &  & \qquad\qquad\qquad\qquad\qquad\:\quad\left.+i\eta_{a}\eta_{b}^{2}s_{1}(c_{2}-1)e^{i\mathbf{k}_{1}\cdot\mathbf{r}-i\omega_{b}\tau-i\omega_{a}T}e^{-iH_{a}T}e^{-iH_{b}\tau}\right]\left|\alpha\beta\chi\right\rangle \nonumber \\
 &  & +\left[i\eta_{b}s_{3}\left|g\right\rangle e^{-i\mathbf{k}_{3}\cdot\mathbf{r}-iH_{g}t}+(c_{3}-1)\eta_{a}\eta_{b}\left|a\right\rangle e^{-i\omega_{a}t-iH_{a}t}+(\eta_{a}^{2}+\eta_{b}^{2}c_{3})\left|b\right\rangle e^{-i\omega_{b}t-iH_{b}t}\right]\label{eq:threepulsewavefunctionwithenvironment}\\
 &  & \quad\otimes\left[i\eta_{b}c_{1}s_{2}e^{i\mathbf{k}_{2}\cdot\mathbf{r}-i\omega_{b}T}e^{-iH_{b}T}e^{-iH_{g}\tau}+i\eta_{a}s_{1}(c_{2}-1)e^{i\mathbf{k}_{1}\cdot\mathbf{r}-i\omega_{a}\tau-i\omega_{b}T}e^{-iH_{b}T}e^{-iH_{a}\tau}\right.\nonumber \\
 &  & \qquad\qquad\qquad\qquad\qquad\:\quad\left.+i\eta_{a}\eta_{b}^{2}s_{1}(\eta_{a}^{2}+\eta_{b}^{2}c_{2})e^{i\mathbf{k}_{1}\cdot\mathbf{r}-i\omega_{b}\left(\tau+T\right)}e^{-iH_{b}\left(\tau+T\right)}\right]\left|\alpha\beta\chi\right\rangle .
\end{eqnarray}

\section{Trajectory dephasing factor\label{sec:Trajectory-dephasing-factor}}

The trajectory factors $\mathcal{D}_{aa}\left(\tau,T;\alpha\beta\chi\right)$
and $\mathcal{D}_{8}\left(\tau,T,t;\alpha,\beta,\chi\right)$ in 2PPE
and 3PPE, respectively, are presented. By computing the evolution
of the coherent state and taking the overlap, we can get

\begin{align}
\mathcal{D}_{aa}\left(\tau,T;\alpha\beta\chi\right) & =\exp\left(-2g_{a}^{0*}\left(\tau\right)-g_{a}^{0*}\left(T\right)-g_{a}^{0}\left(T\right)+g_{a}^{0*}\left(\tau+T\right)\right)\nonumber \\
 & \times\exp\left(-2\mathscr{C}_{a}^{0*}\left(\tau\right)-\mathscr{C}_{a}^{0*}\left(T\right)-\mathscr{C}_{a}^{0}\left(T\right)+\mathscr{C}_{a}^{0*}\left(\tau+T\right)\right)\nonumber \\
 & \times\exp\left\{ i2\left[w\textrm{Im}\left(\alpha\right)\left(1-2\cos\omega\tau+\cos\left(\omega\left(T+\tau\right)\right)\right)+w\textrm{Re}\left(\alpha\right)\left(2\sin\omega\tau-\sin\left(\omega\left(T+\tau\right)\right)\right)\right]\right\} \nonumber \\
 & \times\exp\left\{ i2\left[u_{a}\textrm{Im}\left(\chi\right)\left(1-2\cos\mu\tau+\cos\left(\mu\left(T+\tau\right)\right)\right)+u_{a}\textrm{Re}\left(\chi\right)\left(2\sin\mu\tau-\sin\left(\mu\left(T+\tau\right)\right)\right)\right]\right\} .\label{eq:D_aa_trajectoryfactor}
\end{align}

\begin{align}
\mathcal{D}_{8}\left(\tau,T,t;\alpha,\beta,\chi\right) & =\exp\left[-g_{a}^{0*}\left(T+\tau\right)-g_{b}^{0}\left(t+T\right)\right]\nonumber \\
 & \times\exp\left[-\mathscr{C}_{bb}^{0}\left(t+T\right)-\mathscr{C}_{aa}^{0*}\left(T+\tau\right)-\mathscr{C}_{ab}^{0*}\left(\tau\right)+\mathscr{C}_{ab}^{0*}\left(T+t+\tau\right)+\mathscr{C}_{ab}^{0*}\left(T\right)-\mathscr{C}_{ab}^{0*}\left(t\right)\right]\nonumber \\
 & \exp\left(i2w\left[\textrm{Re}\left(\alpha\right)\sin\left(\omega\left(T+\tau\right)\right)+\textrm{Im}\left(\alpha\right)\left(1-\cos\left(\omega\left(T+\tau\right)\right)\right)\right]\right)\nonumber \\
 & \exp\left(i2v\left[\textrm{Re}\left(\beta\right)\left(\sin\left(\nu\tau\right)-\sin\left(\nu\left(T+\tau+t\right)\right)\right)+\textrm{Im}\left(\beta\right)\left(-\cos\left(\nu\tau\right)+\cos\left(\nu\left(T+\tau+t\right)\right)\right)\right]\right)\nonumber \\
 & \exp\left\{ i2\textrm{Re}\left(\chi\right)\left(u_{b}\sin\left(\mu\tau\right)-u_{b}\sin\left(\mu\left(T+t+\tau\right)\right)+u_{a}\sin\left(\mu\left(T+\tau\right)\right)\right)\right\} \nonumber \\
 & \exp\left\{ i2\textrm{Im}\left(\chi\right)\left(-u_{b}\cos\left(\mu\tau\right)+u_{a}+u_{b}\cos\left(\mu\left(T+t+\tau\right)\right)\right)-u_{a}\cos\left(\mu\left(T+\tau\right)\right)\right\} .\label{eq:D_8_trajectoryfactor}
\end{align}

The other trajectory factors are also calculated in the same way.

\section{Three-pulse photon echo depahsing factor\label{sec:Decoherence-Factor of three pulses}}

Through the simple multivariate Gaussian integration of trajectory
factors the with respect to $\alpha$, $\beta$ and $\chi$ in 3PPE,
the trajectory dephasing factors of each term are given. The trajectory
dephasing factors of the third-order terms are

\begin{align}
D_{1}\left(\tau,T,t\right) & =\exp\left[-g_{a}^{*}\left(\tau\right)-g_{a}^{*}\left(T+\tau\right)+g_{a}^{*}\left(T+t+\tau\right)+g_{a}^{*}\left(T\right)-g_{a}^{*}\left(T+t\right)-g_{a}\left(t\right)\right]\nonumber \\
 & \times\exp\left[-\mathscr{C}_{aa}^{*}\left(\tau\right)-\mathscr{C}_{aa}^{*}\left(T+\tau\right)+\mathscr{C}_{aa}^{*}\left(T+t+\tau\right)+\mathscr{C}_{aa}^{*}\left(T\right)-\mathscr{C}_{aa}^{*}\left(T+t\right)-\mathscr{C}_{aa}\left(t\right)\right],\nonumber \\
D_{2}\left(\tau,T,t\right) & =\exp\left[-g_{b}^{*}\left(\tau\right)-g_{b}^{*}\left(T+\tau\right)+g_{b}^{*}\left(T+t+\tau\right)+g_{b}^{*}\left(T\right)-g_{b}^{*}\left(T+t\right)-g_{b}\left(t\right)\right]\nonumber \\
 & \times\exp\left[-\mathscr{C}_{bb}^{*}\left(\tau\right)-\mathscr{C}_{bb}^{*}\left(T+\tau\right)+\mathscr{C}_{bb}^{*}\left(T+t+\tau\right)+\mathscr{C}_{bb}^{*}\left(T\right)-\mathscr{C}_{bb}^{*}\left(T+t\right)-\mathscr{C}_{bb}\left(t\right)\right],\nonumber \\
D_{3}\left(\tau,T,t\right) & =\exp\left[-g_{a}\left(t\right)-g_{b}^{*}\left(\tau\right)\right]\nonumber \\
 & \times\exp\left[-\mathscr{C}_{aa}\left(t\right)-\mathscr{C}_{bb}^{*}\left(\tau\right)-\mathscr{C}_{ab}^{*}\left(T+\tau\right)+\mathscr{C}_{ab}^{*}\left(T+t+\tau\right)+\mathscr{C}_{ab}^{*}\left(T\right)-\mathscr{C}_{ab}^{*}\left(T+t\right)\right],\nonumber \\
D_{4}\left(\tau,T,t\right) & =\exp\left[-g_{a}^{*}\left(\tau\right)-g_{b}\left(t\right)\right]\nonumber \\
 & \times\exp\left[-\mathscr{C}_{aa}^{*}\left(\tau\right)-\mathscr{C}_{bb}\left(t\right)-\mathscr{C}_{ab}^{*}\left(T+\tau\right)+\mathscr{C}_{ab}^{*}\left(T+t+\tau\right)+\mathscr{C}_{ab}^{*}\left(T\right)-\mathscr{C}_{ab}^{*}\left(T+t\right)\right],\nonumber \\
D_{5}\left(\tau,T,t\right) & =\exp\left[-g_{a}^{*}\left(T+\tau\right)-g_{a}^{*}\left(\tau\right)+g_{a}^{*}\left(T+t+\tau\right)+g_{a}\left(T\right)-g_{a}^{*}\left(t\right)-g_{a}\left(T+t\right)\right]\nonumber \\
 & \times\exp\left[-\mathscr{C}_{aa}^{*}\left(T+\tau\right)-\mathscr{C}_{aa}^{*}\left(\tau\right)+\mathscr{C}_{aa}^{*}\left(T+t+\tau\right)+\mathscr{C}_{aa}\left(T\right)-\mathscr{C}_{aa}^{*}\left(t\right)-\mathscr{C}_{aa}\left(T+t\right)\right],\nonumber \\
D_{6}\left(\tau,T,t\right) & =\exp\left[-g_{b}^{*}\left(T+\tau\right)-g_{b}^{*}\left(\tau\right)+g_{b}^{*}\left(T+t+\tau\right)+g_{b}\left(T\right)-g_{b}^{*}\left(t\right)-g_{b}\left(T+t\right)\right]\nonumber \\
 & \times\exp\left[-\mathscr{C}_{bb}^{*}\left(T+\tau\right)-\mathscr{C}_{bb}^{*}\left(\tau\right)+\mathscr{C}_{bb}^{*}\left(T+t+\tau\right)+\mathscr{C}_{bb}\left(T\right)-\mathscr{C}_{bb}^{*}\left(t\right)-\mathscr{C}_{bb}\left(T+t\right)\right],\nonumber \\
D_{7}\left(\tau,T,t\right) & =\exp\left[-g_{a}\left(T+t\right)-g_{b}^{*}\left(T+\tau\right)\right]\nonumber \\
 & \times\exp\left[-\mathscr{C}_{aa}\left(T+t\right)-\mathscr{C}_{bb}^{*}\left(\tau+T\right)-\mathscr{C}_{ab}^{*}\left(\tau\right)-\mathscr{C}_{ab}^{*}\left(t\right)+\mathscr{C}_{ab}\left(T\right)+\mathscr{C}_{ab}^{*}\left(\tau+T+t\right)\right],\nonumber \\
D_{8}\left(\tau,T,t\right) & =\exp\left[-g_{a}^{*}\left(T+\tau\right)-g_{b}\left(t+T\right)\right]\nonumber \\
 & \times\exp\left[-\mathscr{C}_{bb}\left(t+T\right)-\mathscr{C}_{aa}^{*}\left(T+\tau\right)-\mathscr{C}_{ab}^{*}\left(\tau\right)+\mathscr{C}_{ab}^{*}\left(T+t+\tau\right)+\mathscr{C}_{ab}\left(T\right)-\mathscr{C}_{ab}^{*}\left(t\right)\right].
\end{align}

The trajectory dephasing factors of the fifth-order terms are

\begin{align}
D_{9}\left(\tau,T,t\right) & =\exp\left(-g_{a}^{*}\left(t\right)+g_{a}^{*}\left(T+t\right)-g_{a}\left(T+t\right)-g_{a}^{*}\left(T\right)+g_{a}\left(T\right)-g_{b}^{*}\left(\tau\right)\right)\nonumber \\
 & \times\exp\left(-\mathscr{C}_{aa}^{*}\left(t\right)+\mathscr{C}_{aa}^{*}\left(T+t\right)-\mathscr{C}_{aa}\left(T+t\right)-\mathscr{C}_{aa}^{*}\left(T\right)+\mathscr{C}_{aa}\left(T\right)-\mathscr{C}_{bb}^{*}\left(\tau\right)\right)\nonumber \\
 & \times\exp\left(\mathscr{C}_{ab}^{*}\left(T+t+\tau\right)-\mathscr{C}_{ab}^{*}\left(T+t\right)-\mathscr{C}_{ab}^{*}\left(T+\tau\right)+\mathscr{C}_{ab}^{*}\left(T\right)\right),\nonumber \\
D_{10}\left(\tau,T,t\right) & =\exp\left(-g_{a}^{*}\left(\tau\right)-g_{b}^{*}\left(t\right)+g_{b}^{*}\left(T+t\right)-g_{b}\left(T+t\right)-g_{b}^{*}\left(T\right)+g_{b}\left(T\right)\right)\nonumber \\
 & \times\exp\left(-\mathscr{C}_{aa}^{*}\left(\tau\right)-\mathscr{C}_{bb}^{*}\left(t\right)+\mathscr{C}_{bb}^{*}\left(T+t\right)-\mathscr{C}_{bb}\left(T+t\right)-\mathscr{C}_{bb}^{*}\left(T\right)+\mathscr{C}_{bb}\left(T\right)\right)\nonumber \\
 & \times\exp\left(\mathscr{C}_{ab}^{*}\left(T+t+\tau\right)-\mathscr{C}_{ab}^{*}\left(T+t\right)-\mathscr{C}_{ab}^{*}\left(T+\tau\right)+\mathscr{C}_{ab}^{*}\left(T\right)\right),\nonumber \\
D_{11}\left(\tau,T,t\right) & =\exp\left(-g_{a}^{*}\left(\tau\right)-g_{b}\left(t\right)-\mathscr{C}_{aa}^{*}\left(\tau\right)-\mathscr{C}_{bb}\left(t\right)\right)\nonumber \\
 & \times\exp\left(-\mathscr{C}_{ab}^{*}\left(T+\tau\right)+\mathscr{C}_{ab}^{*}\left(T+\tau+T\right)+\mathscr{C}_{ab}\left(T\right)-\mathscr{C}_{ab}\left(T+t\right)+\mathscr{C}_{ab}\left(t\right)-\mathscr{C}_{ab}^{*}\left(t\right)\right),\nonumber \\
D_{12}\left(\tau,T,t\right) & =\exp\left(-g_{a}\left(t\right)-g_{b}^{*}\left(\tau\right)-\mathscr{C}_{aa}\left(t\right)-\mathscr{C}_{bb}^{*}\left(\tau\right)\right)\nonumber \\
 & \times\exp\left(-\mathscr{C}_{ab}^{*}\left(T+\tau\right)+\mathscr{C}_{ab}^{*}\left(T+\tau+t\right)+\mathscr{C}_{ab}\left(T\right)-\mathscr{C}_{ab}\left(T+t\right)+\mathscr{C}_{ab}\left(t\right)-\mathscr{C}_{ab}^{*}\left(t\right)\right),\nonumber \\
D_{13}\left(\tau,T,t\right) & =\exp\left(-2g_{a}^{*}\left(\tau\right)+g_{a}^{*}\left(T+t+\tau\right)-g_{a}\left(T+t\right)-g_{a}^{*}\left(T+t\right)-g_{b}^{*}\left(T\right)\right)\nonumber \\
 & \times\exp\left(-2\mathscr{C}_{aa}^{*}\left(\tau\right)+\mathscr{C}_{aa}^{*}\left(T+t+\tau\right)-\mathscr{C}_{aa}\left(T+t\right)-\mathscr{C}_{aa}^{*}\left(T+t\right)-\mathscr{C}_{bb}^{*}\left(T\right)\right)\nonumber \\
 & \times\exp\left(\mathscr{C}_{ab}^{*}\left(\tau\right)-\mathscr{C}_{ab}^{*}\left(T+\tau\right)+\mathscr{C}_{ab}^{*}\left(T+t\right)-\mathscr{C}_{ab}^{*}\left(t\right)+\mathscr{C}_{ab}\left(T\right)+\mathscr{C}_{ab}^{*}\left(T\right)\right),\nonumber \\
D_{14}\left(\tau,T,t\right) & =\exp\left(-g_{a}^{*}\left(T\right)-2g_{b}^{*}\left(\tau\right)+g_{b}^{*}\left(T+t+\tau\right)-g_{b}\left(T+t\right)-g_{b}^{*}\left(T+t\right)\right)\nonumber \\
 & \times\exp\left(-\mathscr{C}_{aa}^{*}\left(T\right)-2\mathscr{C}_{bb}^{*}\left(\tau\right)+\mathscr{C}_{bb}^{*}\left(T+t+\tau\right)-\mathscr{C}_{bb}\left(T+t\right)-\mathscr{C}_{bb}^{*}\left(T+t\right)\right)\nonumber \\
 & \times\exp\left(\mathscr{C}_{ab}^{*}\left(\tau\right)-\mathscr{C}_{ab}^{*}\left(T+\tau\right)+\mathscr{C}_{ab}^{*}\left(T+t\right)-\mathscr{C}_{ab}^{*}\left(t\right)+\mathscr{C}_{ab}\left(T\right)+\mathscr{C}_{ab}^{*}\left(T\right)\right)\nonumber \\
D_{15}\left(\tau,T,t\right) & =\exp\left(-g_{a}\left(T\right)-g_{b}\left(t\right)-g_{b}^{*}\left(t\right)+g_{b}^{*}\left(T+\tau+t\right)-2g_{b}^{*}\left(T+\tau\right)\right)\nonumber \\
 & \times\exp\left(-\mathscr{C}_{aa}\left(T\right)-\mathscr{C}_{bb}\left(t\right)-\mathscr{C}_{bb}^{*}\left(t\right)+\mathscr{C}_{bb}^{*}\left(T+\tau+t\right)-2\mathscr{C}_{bb}^{*}\left(T+\tau\right)\right)\nonumber \\
 & \times\exp\left(\mathscr{C}_{ab}^{*}\left(T+\tau\right)-\mathscr{C}_{ab}^{*}\left(\tau\right)+2\mathscr{C}_{ab}\left(T\right)+\mathscr{C}_{ab}\left(t\right)-\mathscr{C}_{ab}\left(T+t\right)\right)\nonumber \\
D_{16}\left(\tau,T,t\right) & =\exp\left(-g_{b}\left(T\right)-g_{a}\left(t\right)-g_{a}^{*}\left(t\right)+g_{a}^{*}\left(T+\tau+t\right)-2g_{a}^{*}\left(T+\tau\right)\right)\nonumber \\
 & \times\exp\left(-\mathscr{C}_{bb}\left(T\right)-\mathscr{C}_{aa}\left(t\right)-\mathscr{C}_{aa}^{*}\left(t\right)+\mathscr{C}_{aa}^{*}\left(T+\tau+t\right)-2\mathscr{C}_{aa}^{*}\left(T+\tau\right)\right)\nonumber \\
 & \times\exp\left(\mathscr{C}_{ab}^{*}\left(T+\tau\right)-\mathscr{C}_{ab}^{*}\left(\tau\right)+2\mathscr{C}_{ab}\left(T\right)+\mathscr{C}_{ab}\left(t\right)-\mathscr{C}_{ab}\left(T+t\right)\right).
\end{align}

The trajectory dephasing factors of the seventh-order terms are

\begin{align}
D_{17}\left(\tau,T,t\right) & =\exp\left(-g_{a}^{*}\left(\tau\right)-g_{a}^{*}\left(T+\tau\right)+g_{a}^{*}\left(T+\tau+t\right)+g_{a}^{*}\left(T\right)-g_{a}^{*}\left(T+t\right)-g_{a}\left(t\right)\right)\nonumber \\
 & \times\exp\left(-\mathscr{C}_{aa}^{*}\left(\tau\right)-\mathscr{C}_{aa}^{*}\left(T+\tau\right)+\mathscr{C}_{aa}^{*}\left(T+\tau+t\right)+\mathscr{C}_{aa}^{*}\left(T\right)-\mathscr{C}_{aa}^{*}\left(T+t\right)-\mathscr{C}_{aa}\left(t\right)\right),\nonumber \\
D_{18}\left(\tau,T,t\right) & =\exp\left(-g_{b}^{*}\left(\tau\right)-g_{b}^{*}\left(T+\tau\right)+g_{b}^{*}\left(T+\tau+t\right)+g_{b}^{*}\left(T\right)-g_{b}^{*}\left(T+t\right)-g_{b}\left(t\right)\right)\nonumber \\
 & \times\exp\left(-\mathscr{C}_{bb}^{*}\left(\tau\right)-\mathscr{C}_{bb}^{*}\left(T+\tau\right)+\mathscr{C}_{bb}^{*}\left(T+\tau+t\right)+\mathscr{C}_{bb}^{*}\left(T\right)-\mathscr{C}_{bb}^{*}\left(T+t\right)-\mathscr{C}_{bb}\left(t\right)\right),\nonumber \\
D_{19}\left(\tau,T,t\right) & =\exp\left(-2g_{a}^{*}\left(\tau\right)-g_{a}^{*}\left(T\right)-g_{a}\left(T\right)+g_{a}^{*}\left(T+\tau\right)-2g_{b}^{*}\left(T\right)-g_{b}^{*}\left(t\right)-g_{b}\left(t\right)+g_{b}^{*}\left(T+t\right)\right)\nonumber \\
 & \times\exp\left(-2\mathscr{C}_{aa}^{*}\left(\tau\right)-\mathscr{C}_{aa}^{*}\left(T\right)-\mathscr{C}_{aa}\left(T\right)+\mathscr{C}_{aa}^{*}\left(T+\tau\right)-2\mathscr{C}_{bb}^{*}\left(T\right)-\mathscr{C}_{bb}^{*}\left(t\right)-\mathscr{C}_{bb}\left(t\right)+\mathscr{C}_{bb}^{*}\left(T+t\right)\right)\nonumber \\
 & \times\exp\left(\mathscr{C}_{ab}^{*}\left(\tau\right)-2\mathscr{C}_{ab}^{*}\left(T+\tau\right)+\mathscr{C}_{ab}^{*}\left(T+\tau+t\right)+3\mathscr{C}_{ab}^{*}\left(T\right)+2\mathscr{C}_{ab}\left(T\right)-\mathscr{C}_{ab}\left(T+t\right)-\mathscr{C}_{ab}^{*}\left(T+t\right)+\mathscr{C}_{ab}\left(t\right)\right),\nonumber \\
D_{20}\left(\tau,T,t\right) & =\exp\left(-2g_{b}^{*}\left(\tau\right)-g_{b}^{*}\left(T\right)-g_{b}\left(T\right)+g_{b}^{*}\left(T+\tau\right)-2g_{a}^{*}\left(T\right)-g_{a}^{*}\left(t\right)-g_{a}\left(t\right)+g_{a}^{*}\left(T+t\right)\right)\nonumber \\
 & \times\exp\left(-2\mathscr{C}_{bb}^{*}\left(\tau\right)-\mathscr{C}_{bb}^{*}\left(T\right)-\mathscr{C}_{bb}\left(T\right)+\mathscr{C}_{bb}^{*}\left(T+\tau\right)-2\mathscr{C}_{aa}^{*}\left(T\right)-\mathscr{C}_{aa}^{*}\left(t\right)-\mathscr{C}_{aa}\left(t\right)+\mathscr{C}_{aa}^{*}\left(T+t\right)\right)\nonumber \\
 & \times\exp\left(\mathscr{C}_{ab}^{*}\left(\tau\right)-2\mathscr{C}_{ab}^{*}\left(T+\tau\right)+\mathscr{C}_{ab}^{*}\left(T+\tau+t\right)+3\mathscr{C}_{ab}^{*}\left(T\right)+2\mathscr{C}_{ab}\left(T\right)-\mathscr{C}_{ab}\left(T+t\right)-\mathscr{C}_{ab}^{*}\left(T+t\right)+\mathscr{C}_{ab}\left(t\right)\right).
\end{align}

\end{widetext}
\end{document}